\documentclass[pra,aps,final,twocolumn,nofootinbib,showpacs]{revtex4-1}
\usepackage{epsfig}
\usepackage{latexsym}
\usepackage{xspace}
\usepackage{hyperref}
\usepackage[latin2]{inputenc}
\usepackage{indentfirst}
\usepackage{enumerate}
\usepackage{color}
\usepackage{colordvi}

\usepackage{amsmath}
\usepackage{amssymb}
\usepackage[english]{babel}
\usepackage{url}

\def\be{\begin{equation}}
\def\ee{\end{equation}}
\def\bea{\begin{eqnarray}}
\def\eea{\end{eqnarray}}
\def\l{\label}

\def\p{{\bf p}}

\def\q{{\bf q}}

\def\d{\mbox{d}}

\def\siml{\;\hbox{\kern.1em \lower.7ex \hbox{$\sim$} \kern-1.12em
 \raise.5ex \hbox{$<$} \kern.1em}}
\def\simg{\;\hbox{\kern.1em \lower.7ex \hbox{$\sim$} \kern-1.12em
 \raise.5ex \hbox{$>$} \kern.1em}}

\def\siml{\;\hbox{\kern.1em \lower.7ex \hbox{$\sim$} \kern-1.12em
 \raise.5ex \hbox{$<$} \kern.1em}}
\def\simg{\;\hbox{\kern.1em \lower.7ex \hbox{$\sim$} \kern-1.12em
 \raise.5ex \hbox{$>$} \kern.1em}}

\begin{document}

\title{Quantum statistics effects and fluctuations of particle numbers
  near the critical point of nuclear matter
}

\author{ S.N.\ Fedotkin, A.G.\ Magner, and U.V.\ Grygoriev}
\affiliation{Institute for Nuclear Research NASU, 03680 Kiev, Ukraine}

\begin{abstract}
Equation of state with 
quantum statistics corrections
is derived for a multi-component gas of particles interacting through the 
repulsive and attractive van der Waals (vdW) forces 
up to first few orders
 over a small  parameter
  $\delta \approx \hbar^3 n(mT)^{-3/2}[g(1- bn)]^{-1}$, where
$n$ and $T$ are the particle number density and temperature,
$m$ and $g$ the particle mass and
 degeneracy factor.
 The parameter $b$ corresponds to the 
 vdW excluded volume. 
    For interacting system of Fermi nucleon and Bose
  $\alpha$  particles, a small impurity of $\alpha$
  particles to the nucleon system at leading first order in both
  $\alpha $ particle and nucleon small parameters $\delta $
  does not change much the basic results for
  the symmetric nuclear matter.
 The particle number fluctuations $\omega$ determined by 
 the isothermal in-compressibility $\mathcal{K}(n,T)$ can be obtained
 analytically at the same first order quantum-statistics approximation
 for symmetric nucleon matter.
 Our approximate analytical results appear to be in 
  good
  agreement with the
  accurate numerical calculations.
  
\end{abstract}
%

\maketitle

\section{Introduction}
A study of 
hadron matter, first of all, an interacting system of protons and neutrons,
has a long history; see, e.g.,
Refs.~\cite{nm-1,nm-2,nm-3,nm-4,nm-5,nm-6,nm-7,nm-8,nm-9,nm-10,nm-11}.
Realistic versions of the nuclear matter equation of state includes both
the attractive and repulsive forces
between
particles. 
Thermodynamical behavior of this matter
leads to the liquid-gas first-order phase transition
which ends at the critical point.
Experimentally, a presence of the liquid-gas phase transition
in nuclear matter was  reported and then analyzed
in numerous papers
(see, e.g., Refs.~\cite{ex-1,ex-2,ex-3,ex-4,ex-5,ex-6}).
Critical points in different systems of hadrons were studied in
Refs.~\cite{vova,satarov,roma,roma1}, see also references therein.

Recently, the proposed  van der Waals (vdW) equation of state
accounting for 
the quantum statistics  (QS)~\cite{marik,vova,FMG19}
was used to describe the
properties of 
hadronic matter, also  with many component extensions and applications to the
fluctuation calculations for different thermodynamical averages
\cite{TR38,RJ58,LLv5,TK66,IA71,BR75,AC90,ZM02}. 
The role and size of the effects of 
QS was studied analytically for nuclear matter, also
for pure neutron and pure $\alpha$-particle matter in Ref.~\cite{FMG19}.
Particularly,
we investigated a dependence
of the critical point parameters on
the particle mass $m$,
degeneracy factor $g$, and the 
vdW parameters $a$ and $b$
which describe particle interactions for 
each of these systems.
Our consideration 
was restricted to small temperatures, 
$T \siml 30$~MeV, and not too large
particle densities.  Within these restrictions, the
number of nucleons
becomes a conserved number,
and the chemical potential of 
such systems regulates the number density of 
particles. An extension to the fully relativistic hadron resonances in a gas
formulation with 
vdW interactions between
baryons and between antibaryons was considered in 
Ref.~\cite{VGS-17}.
An application of this extended model to net baryon number fluctuations
in relativistic nucleus-nucleus
collisions was developed in Ref.~\cite{VJGS-18}.
We do not include the Coulomb forces
and make no differences between protons and neutrons
(both these particles are named as nucleons).
In addition, under these restrictions the
non-relativistic treatment becomes very accurate
and is adopted in our studies. In the present work we are going to apply
the same analytical method as in Ref.~\cite{FMG19} to the
mixed two-component
system of nucleons and $\alpha$ particles. Another attractive subject of
this work is to apply our analytical results to analysis of the
particle number fluctuations near the critical points of
the nuclear matter.

The paper is organized as the following. In Sec.~\ref{sec-2} we
recall some results of the
ideal Bose and Fermi gases taking an exemplary case of the
two-component $N-\alpha$ system.
In Sec.~\ref{sec-3} the QS 
effects
near the critical point are studied for the system of symmetric-nuclear
and $\alpha$-particle matter. Our analytical results are used for nucleon number
fluctuations in Sec.~\ref{sec-4}.
These results are then discussed in Sec.~\ref{sec-5}
and summarized in
Sec.~\ref{sec-6}.

\section{Ideal quantum gases}\label{sec-2}

The pressure $P_i(T,\mu)$ for the $i$-system of particles (e.g.,
$i=\{N,\alpha\}$)
plays the role of the thermodynamical potential in
the grand canonical ensemble (GCE)
where temperature
$T$ and chemical potential $\mu$ are independent variables.
The particle number density
$n_i(T,\mu)$, entropy density $s_i(T,\mu)$, and energy density
$\varepsilon_i(T,\mu)$  are given as
\be\label{term}
  n_i=\left(\frac{\partial P_i}{\partial \mu}\right)_T~,~
  s_i=\left(\frac{\partial P_i}{\partial T}\right)_\mu~,~
\varepsilon_i= Ts_i+\mu n_i-P_i~.
\ee
In the thermodynamic limit $V\rightarrow \infty$ considered in the present
paper all intensive thermodynamical  functions -- $P$, $n$, $s$,
and $\varepsilon$ --
depend on $T$ and $\mu$, 
 rather than on the system volume $V$, see for instance
 Ref.~\cite{BG-08}.
 We start with the
GCE expressions $\sum_iP^{\rm id}_i(T,\mu)$ for the pressure $P^{\rm id}(T,\mu)$
and particle number density $n^{\rm id}(T,\mu)=\sum_in^{\rm id}_i(T,\mu)$
 for the ideal
non-relativistic
quantum gas \cite{G,LLv5},
\bea\l{Pid}
& P^{\rm id}_i=\frac13 g_i\int \frac{d {\bf p}}{(2\pi \hbar)^3}\frac{p^2}{m_i}
  \left[\exp\left( \frac{p^2}{2m_iT} -\frac{\mu}{T }\right) -
    \theta_i\right]^{-1},\\
& n^{\rm id}_i=g_i\int \frac{d {\bf p}}{(2\pi \hbar)^3}
  \left[\exp \left( \frac{p^2}{2m_iT} -\frac{\mu}{T} \right) -
    \theta_i\right]^{-1}~,\l{nid}
\eea
where $m_i$ and $g_i$
are, respectively, the particle mass and degeneracy factor of the $i$
component. The value of $\theta_i=-1$ corresponds to the Fermi gas,
$\theta_i=1$ to the Bose gas, and $\theta_i=0$ is the Boltzmann (classical)
approximation when
effects of the
QS are neglected\footnote{The units
  with Boltzmann  constant
    $\kappa^{}_{\rm B}=1$ are used. We keep the Plank constant in the
  formulae  to illustrate
  the effects of QS, 
  but put $\hbar=h/2\pi=1$ in all
  numerical calculations.  For simplicity, we omitted here and below the
 subscript id for the ideal gas everywhere where
it will not lead to a misunderstanding. }.

Equations (\ref{Pid}) and (\ref{nid}) can be expressed in terms of the
power series over fugacity, $z \equiv \exp(\mu/T)$, 
as:
\bea
 P_i(T,z) &
   \equiv  \frac{g_iT}{\theta_i \Lambda^3_i}\,{\rm Li}_{5/2}(\theta_i z) =
  \frac{g_iT}{\theta_i \Lambda^3_i}\,\sum_{k=1}^\infty
  \frac{(\theta_i z)^k}{k^{5/2}}~ ,\l{Pid-1}\\ 
  n_i(T,z) &
  \equiv
    \frac{g_i}{\theta_i \Lambda^3_i}\,{\rm Li}_{3/2}(\theta z)
    =  \frac{g_i}{\theta_i \Lambda^3_i}\,\sum_{k=1}^\infty
    \frac{(\theta_i z)^k}{k^{3/2}}~ .
    \l{nid-1}
    \eea
    Here,
\be\l{lambdaT}
\Lambda_i~\equiv ~\hbar\sqrt\frac{2 \pi}{m_iT}~
\ee
is the de Broglie 
thermal wavelength  \cite{LLv5},
and $\mbox{Li}_\nu$
is the polylogarithmic function \cite{Grad-Ryzhik,Li}.
The values of $\mu >0$, i.e., $z>1$, are forbidden
in the ideal Bose  gas. The point
$\mu=0$ corresponds to an onset of the Bose-Einstein condensation in the system
of bosons. For fermions, any values of $\mu$
are possible, i.e., integrals (\ref{Pid}) and (\ref{nid}) exist for
$\theta_i=-1$ at all real values of $\mu$.
The power series
(\ref{Pid-1}) and (\ref{nid-1}) are obviously convergent at
$z < 1$. For the Fermi statistics at $z>1$, the integral
representation of the  corresponding polylogarithmic function
can be used.
Particularly,
at $z\rightarrow \infty$ one can use the asymptotic Sommerfeld expansion of the
$\mbox{Li}_{\nu}(-z)$ functions over $1/\mbox{ln}^2|z|$
\cite{brack}.

For nucleon gas we take $m^{}_N \cong 938$~MeV neglecting
a small difference between proton and neutron masses.
The degeneracy factor is then  $g^{}_N=4$ which takes into account
two spin and two isospin
states of nucleon. 
For ideal Bose gas of $\alpha$-nuclei, 
one has $g_\alpha=1$ and $m_\alpha\cong 3727$~MeV.

At $z\ll 1$,  only one
term $k=1$ is enough  in Eqs.~(\ref{Pid-1}) and (\ref{nid-1}) which leads
to the classical ideal gas relation
\be\l{Pid-cl}
P=n\,T~.
\ee
Note that the result (\ref{Pid-cl}) follows automatically from
Eqs.~(\ref{Pid}) and (\ref{nid})
at $\theta_i=0$.
The classical Boltzmann approximation
at $z\ll 1$
is valid for large $T$  and/or 
small $n$ region of the $n$-$T$ plane.
In fact, at very small $n$, one observes $z<1$ at small $T$ too.

\begin{figure*}
\begin{center}
    \includegraphics[width=10.0cm,clip]{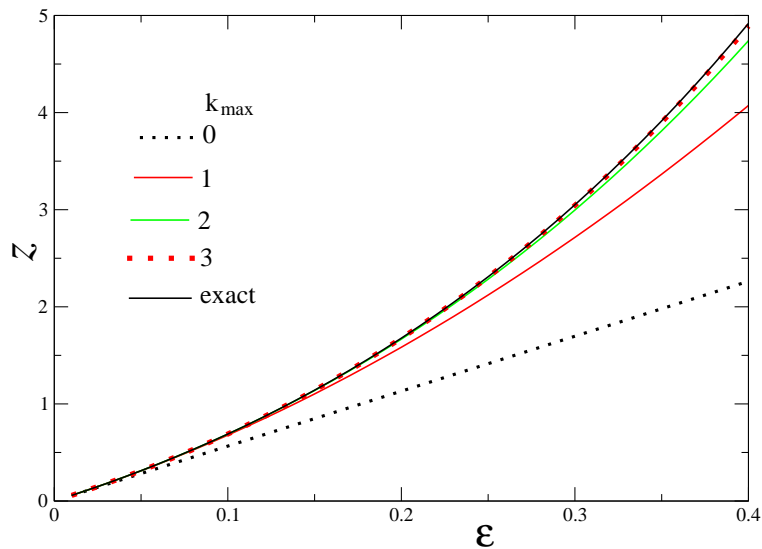}
\end{center}

\caption{
  Fugacity $z$ as function of the
  quantum statistics parameter $\epsilon$
  for small values where one finds the critical points  ($\epsilon_c =0.1-0.2$
  in nuclear matter). Solid black curve shows the exact fugacity $z(\epsilon)$, and
  $k_{\rm max}$ is the maximal power of cut-off series for polylogarithm $Li$.
}
\label{fig1-fe}
\end{figure*}
\begin{figure*}
\begin{center}
  \includegraphics[width=8.0cm,clip]{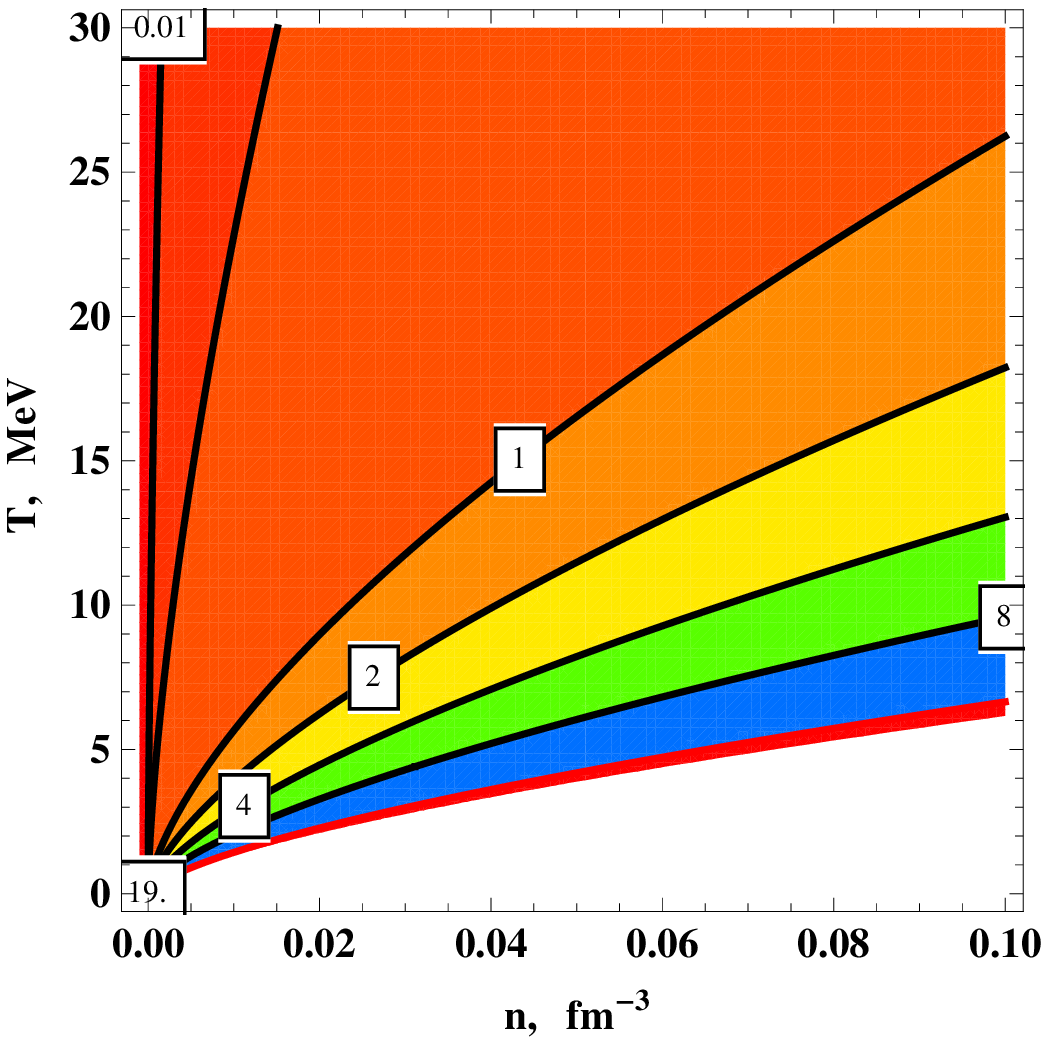}
  \includegraphics[width=8.1cm,clip]{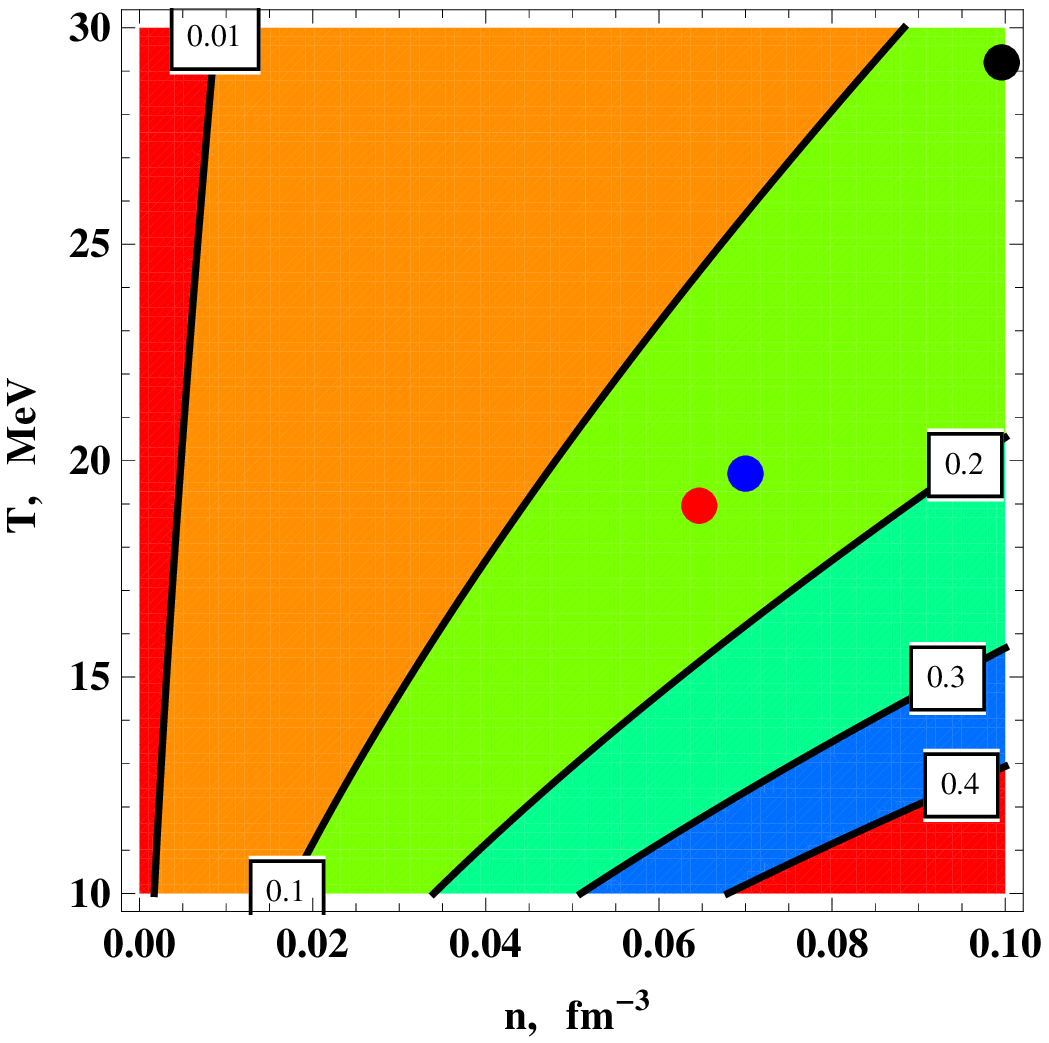}
\end{center}

\caption{
  Contour plots for the first-order fugacity $z(n,T)$
  and parameter $\epsilon(n,T)$ 
  for  nucleon matter in the plane of density
  $n$ and temperature $T$ 
  are shown in 
  left and right panels, respectively.
  The red line 
  (left) shows the zero entropy line, such that the white area is
  related to a nonphysical region where the entropy of the ideal gas
  is
  negative.
The critical point for our first-order and the zero-order (standard vdW)
approximations
for nuclear matter at the parameters $a$ and $b$ [\ref{ab}] are shown on right by
the red and black points, relatively. The blue point in the same plot
presents the numerical result for the critical point 
(Ref.~\cite{FMG19}).
}
\label{fig2}
\end{figure*}

Inverting the $z^k$ power series in  Eq.~(\ref{nid-1}), one 
transforms the power expansion of $z(e_i)$ 
to the parameter $e_i$
(see, e.g., Ref.~\cite{BR75}), 
\be\l{eps}
  e_i = -\frac{\theta_i n_i \Lambda^{3}_i}{4\sqrt2\, g_i}\equiv
  -\theta_i\, \epsilon^{}_{i}~,
  \ee
  where
  \be\l{eps0}
   \epsilon^{}_{i}=
 \, \frac{ \hbar^3\,\pi^{3/2}~n_i}{2\,g_i\,(m_iT)^{3/2}}~
\ee
Taking a given component $i$, e.g., for nucleon matter ($\theta_i=-1$),
for simplicity, we will omit
subscript $i$ in
discussions of Fig.~\ref{fig1-fe}.
The exact fugacity $z(\epsilon)$
can be obtained by multiplying  equation 
(\ref{nid-1}) by the
 factor 
 $\Lambda^3/(4\sqrt{2}~g)$
 to get $\epsilon=\epsilon(z) $ and, then, inverting
  this equation with respect to $z$. Different other curves in
  Fig.~\ref{fig1-fe} present the
  maximal power $k_{\rm max}$ of
  the sum 
  of Eq.~(\ref{nid-1}) over $k$ after the same multiplying and cut-off the
  series 
  for the polylogarithmic function
  $\mbox{Li}(-z)$
  in powers of $z$ (at the order $k_{\rm max}$). As seen from this figure, one has
  the asymptotic convergence  over $k_{\rm max}$ - the better the
    smaller $\epsilon$.
  Even the first-order correction is leading and good in the region
  of $\epsilon\approx \epsilon_c=0.1-0.2$ where $z \approx 1$.
  The second ($k_{\rm max}=2$) correction improves the convergence such that
  the cut-off sum for $\mbox{Li}$ at the power $k_{\rm max}$ practically coincides
  with the exact result (Fig.\ref{fig1-fe}). For larger
  $\epsilon$, say, $\epsilon > 1$, where the fugacity $z$ is much larger
  than 1 (e.g., in the
  small temperature limit), we need more
  and more terms and one has a divergence of the series in $k_{\rm max}$.
  In this region the series for $\mbox{Li}$ 
  fails, and one has to use
  another asymptotic expansion, for instance,  over $1/z^2$ as suggested by
  Zommerfeld \cite{brack}.

  Fig.~\ref{fig2} shows  the contour graphics in the $n-T$ plane where black lines  mean
  $z(n,T)=const$ on left, and $\epsilon(n,T)=const$ on right with the values written in
  white squares. As seens from these plots, all values of $z \siml 1$ correspond to
  $\epsilon \ll 1$ above blue regions, and therefore, together with Fig.~\ref{fig1-fe},
  this explains reasons for using the expansion in small parameter $\epsilon$, even when
  the fugacity is of the order of 1 and somewhat larger. In particular, the critical points
  obtained in Ref.~\cite{FMG19} belong to such a region.

  The expansion of $z(\epsilon)$ in powers of $\epsilon$ is inserted
  then into Eq.~(\ref{Pid-1}).
At small $\epsilon_i <1$ the expansion of the pressure over
the powers of $\epsilon_i$
is rapidly
convergent asymptotically, i.e., converges to the exact
(polylogarithmic) function result (\ref{Pid-1}) and (\ref{nid-1}),
the faster the smaller $\epsilon_i$, such that
a few first terms give already a good approximation of the
QS
effects. Notice that the fugacity values of $z$ can be
larger 1, however,  for small $\epsilon$ and, similarly, for other
corrections of a maximal power $k_{\rm max}$ in the $\mbox{Li}(z)$ polynomials.
Taking   the two 
terms, $k=1$, and $2$, 
in
Eqs.~(\ref{Pid-1}) and (\ref{nid-1}),
one obtains a classical gas result (\ref{Pid-cl})
plus the leading first few-order corrections due to the effects
of QS:
\be\label{Pid-n}
P_i(T,n_i)= n_i T \left[1 +
  e_i - c_2 e_i^2 - c_3 e_i^3 
 + \mbox{O}(e^4_i)\right]~,    
\ee
where
$c^{}_2=4[16/(9\sqrt{3})-1] \cong 0.106$~,
$c_3=4(15+9\sqrt{2}-16\sqrt{3})/3\cong 0.0201$~,
and so on.
For brevity, we call the linear and quadratic
$\epsilon_i$-terms in
  Eq.~(\ref{Pid-n})
  as the first and second
  (order) quantum corrections.

  Equation (\ref{Pid-n}) demonstrates explicitly a deviation of  the quantum
  ideal gas pressure
from the classical ideal-gas  value (\ref{Pid-cl}): the Fermi statistics leads to an increasing of the classical pressure,
while the Bose statistics to its decreasing. This is often interpreted \cite{LLv5} as the effective
Fermi `repulsion' 
and Bose `attraction'
between 
QS particles.

\section{
vdW  model with quantum-statistics corrections}\label{sec-3}

For the infinite system of a mixture of of different particles, e.g.,
Fermi and Bose particles
-- nucleons and $\alpha$ particles, one can present
the pressure function of the vdW model (vdWM) with the QS
(QvdWM) 
\cite{vova}
\bea\l{PQvdW}
&P(T,n)=P^{\rm id}_N(T,\mu^\ast_N) 
+ P^{\rm id}_\alpha(T,\mu^\ast_\alpha) 
\nonumber\\
&-a^{}_{NN} n^2_N -2 a^{}_{N\alpha}n^{}_Nn^{}_\alpha -
a^{}_{\alpha\alpha}n^2_\alpha~,
\eea
where
\bea\l{PNal}
&
P^{\rm id}_N(T,\mu^\ast_N)=
\frac{4g^{}_N~T}{3\sqrt{\pi}~\Lambda^3_N}
\int_0^\infty d\eta
\frac{\eta^{3/2}}{\exp\left(\eta - \frac{\mu^\ast_N}{T}\right) +1}~,\nonumber\\
&
P^{\rm id}_\alpha(T,\mu^\ast_\alpha)=
\frac{4g_\alpha~T}{3\sqrt{\pi}~\Lambda^3_\alpha}
\int_0^\infty d\eta
\frac{\eta^{3/2}}{\exp\left(\eta - \frac{\mu^\ast_\alpha}{T}\right) - 1}~,
\eea
Here $n$ is the baryon number density,
$n=n^{}_N+4 n_{\alpha}$,  $P^{\rm id}_i$ 
is given by
Eq.~(\ref{Pid}), 
and
$\mu^\ast_i$ are the solutions of transcendental  equations:
\bea\l{nNals}
&n^\ast_N = n^{\rm id}_N(T,\mu^\ast_N) \equiv
\frac{2g^{}_N}{\sqrt{\pi}~\Lambda^3_N}
\int_0^\infty d\eta
\frac{\eta^{1/2}}{\exp\left(\eta - \frac{\mu^\ast_N}{T}\right) +1}~,\nonumber\\
&n^\ast_\alpha = n^{\rm id}_\alpha(T,\mu^\ast_\alpha)\equiv
\frac{2g_\alpha~}{\sqrt{\pi}~\Lambda^3_\alpha}
\int_0^\infty d\eta
\frac{\eta^{1/2}}{\exp\left(\eta - \frac{\mu^\ast_\alpha}{T}\right) - 1}~,
\eea
and $n^{\rm id}_i$ is defined by Eq.~(\ref{nid}),
see more details in Refs.~\cite{marik,vova}.
The relationship between the densities $n_i$ of Eq.~(\ref{PQvdW}) and
auxiliary ones $n^\ast_i$ can be written in the following form \cite{vova}:
\bea\l{nN}
&n^{}_N=\frac{n^\ast_N\left[
    1+\left(b_{\alpha\alpha}-b^{}_{\alpha N}\right)n^\ast_\alpha\right]}{
  1+b^{}_{NN}n^\ast_N+b_{\alpha\alpha}n^\ast_\alpha +\left(b^{}_{NN}b_{\alpha\alpha}
-b^{}_{N\alpha}b^{}_{\alpha N}\right)n^\ast_Nn^\ast_\alpha}~,\\
  &n^{}_\alpha=\frac{n^\ast_\alpha\left[
    1+\left(b^{}_{NN}-b^{}_{N \alpha}\right)n^\ast_N\right]}{
  1+b^{}_{NN}n^\ast_N+b_{\alpha\alpha}n^\ast_\alpha +\left(b^{}_{NN}b_{\alpha\alpha}
  -b^{}_{N\alpha}b^{}_{\alpha N}\right)n^\ast_Nn^\ast_\alpha}~,
  \l{nNal}
 \eea
 where $b_{ij}$ are the vdWM exclusion volume constants \cite{vova}:
 \bea\l{bij}
 &b^{}_{NN}=3.35~\mbox{fm}^3,~~~ b^{}_{\alpha\alpha}=16.76~\mbox{fm}^3, \nonumber\\
 &b^{}_{\alpha N}=13.95~\mbox{fm}^3,~~~ b^{}_{N\alpha}=2.85~\mbox{fm}^3~.
 \eea
    Notice that for Bose particles, the restriction $\mu^\ast_\alpha\le 0$
    for the non-relativistic chemical potential
    should be satisfied.
    These restrictions
    correspond to those $\mu_\alpha\le 0$ 
    in the non-relativistic case
    in the ideal Bose gas.
    In Eq.~(\ref{PQvdW}), the  constants
    $a_{ij}>0$ and $b_{ij}>0$ are responsible for respectively
    attractive and repulsive
interactions between particles.\\

In the Boltzmann approximation, i.e. at $\theta=0$ in Eqs.~(\ref{Pid}) and (\ref{nid}), the QvdWM
is reduced to the {\it classical} vdWM 
\cite{LLv5},
\be\l{vdW}
P_i=\sum_{j}\left[\frac{n_iT}{1-n_jb_{ij}} - a_{ij}n_in_j\right]~.
\ee
Note that the classical 
vdWM (\ref{vdW}) is further reduced
to the ideal
classical gas (\ref{Pid-cl}) at $a_{ij}=0$ and $b_{ij}=0$.
At $a_{ij}=0$ and $b_{ij}=0$ the QvdWM
turns into the  {\it quantum} ideal gas
Eqs.~(\ref{Pid}) and (\ref{nid}).

Following Ref.~\cite{vova}, one can fix the model parameters $a_{ij}$ and $b_{ij}$
using the ground state
properties  of the corresponding system components,
(see, e.g., Ref.~\cite{nm}) by 
\be\l{ab}
a^{}_{NN}=329.8\, \mbox{MeV} \cdot \mbox{fm}^3 ~,~~~
a^{}_{N\alpha}=a^{}_{\alpha N}=a_{\alpha\alpha}=0~.
\ee
These values are very close to those
found in  Refs.~\cite{marik,vova}.  Other constants are taken from
Ref.~\cite{vova} [see Eq.~(\ref{bij})].
Small differences appear  because of  the
non-relativistic formulation used in the
present studies. Notice that the system of $N+\alpha$ was studied
in Ref.~\cite{satarov} in the Skyrme model, and the QvdW approach is
criticized because the Bose condensation cannot be described in the
QvdW model.

In what follows, a few 
first 
quantum corrections
of the 
QvdWM
will be considered.
Expanding 
$P^{\rm id}_i\left(T, \mu^\ast_i\right)$, Eq.~(\ref{PNal})
used in  Eq.~(\ref{PQvdW}),
over 
small parameters $\epsilon^\ast_i$ (Eq.~(\ref{eps}) with $i=N, n_i= n^\ast_N $ or
$i=\alpha,n_i= n^\ast_\alpha$, and superscript in $\epsilon^\ast_i$
corresponds to that of
$n^\ast_i$),
one obtains
\be\l{Pvdw-n}
 P^{\rm id}_N(T,n^\ast_N) = n^\ast_N\,T\,\left[1+\epsilon^\ast_{N}\right]
\ee
and
\be\l{Pvdw-a}
 P^{\rm id}_\alpha(T,n^\ast_\alpha) =  n^\ast_\alpha\,T\,\left[1-\epsilon^\ast_{\alpha}\right]
 ~,
\ee
where $\epsilon^\ast_{i}$ is given by Eq.~(\ref{eps0}) with
    replacing
$n_i$ by $n_i^\ast$.
These expressions are similar to those of 
    Eq.~(\ref{Pid-n}) at the first order in $\epsilon_i$.
We proved that  at small $\epsilon^\ast_i$ 
the expansion of the pressure
over powers of $\epsilon^\ast_i$ 
becomes rapidly convergent
    to the exact results, and a  few first
  terms give already a good approximation.
  Our Eq.~(\ref{Pvdw-n}),
     in contrast to Eq.~(\ref{Pid-n})
     discussed in Refs.~\cite{LLv5,BR75},
     takes into account the particle  interaction effects
     (cf. with the previous section \ref{sec-2}).
      A new point of
our consideration is the analytical estimates
of the QS effects
in a mixed system of
interacting fermions and bosons.
     Similarly to the ideal gases, the quantum corrections in
     Eq.~(\ref{Pvdw-n}) increases
     with the particle number density $n_i$ and decreases with the system
     temperature $T$,
     particle mass $m_i$, and degeneracy factor $g_i$.

As in Ref.~\cite{vova}, we introduce now the impurity contribution
of the $\alpha$- particles in the symmetric nuclear matter
as the ratio of the number
of nucleons in the
$\alpha$ particle impurity referred to the total number of nucleons,
\be\l{Xal}
X_\alpha= \frac{4 n_\alpha}{n_N +4 n_\alpha} \equiv
\frac{4 n_\alpha}{n}~,
\ee
where $n$ is the baryon number density defined already above (below
Eq.~(\ref{PNal}).
 According to the numerical solutions in Ref.~\cite{vova},
for the parameters of Eq.~(\ref{ab}), the value of $X_\alpha$ has been
approximately obtained,
$X_\alpha\approx 0.013$. We will use below this value in our calculations.
Taking this
estimate for a simple exemplary case,
one can find $n^\ast_N$ and
$n^\ast_\alpha$ from equations
(\ref{nN}) and (\ref{nNal}). 
 Then, using Eqs.~(\ref{Xal})
and (\ref{bij}), one can present them in the following approximate form:
\be\l{nistar}
n^\ast_N\approx\frac{r^{}_1 n}{1-\tilde{b}^{}_Nn},\quad
n^\ast_\alpha\approx\frac{r^{}_2 n}{1-\tilde{b}^{}_\alpha n}~,
\ee
where
\be\l{tbest}
r^{}_1=(1-X_\alpha)= 0.987, \quad r^{}_2= \frac{X_\alpha}{4}= 0.0033~.
\ee
 Here, 
$\tilde{b}_{i}$  are coefficients related approximately to the
interaction
constants $b_{ij}$, Eq.~(\ref{bij}),
\be\l{tbest1}
\tilde{b}^{}_N\approx 3.29\,\mbox{fm}^3~,
\quad \tilde{b}^{}_\alpha\approx 2.81 \,\mbox{fm}^3~.
\ee
For another interaction  parameter  $a^{}_1$, one can use
\be\l{tbest2}
 a^{}_1=r^2_1\, a^{}_{NN}\approx 321.3 \,\mbox{MeV}\cdot \mbox{fm}^3~.
\ee
 Using also Eq.~(\ref{PQvdW}) with Eqs.~(\ref{Pvdw-n})
    and (\ref{Pvdw-a}), for the parameter values of the
    order of mentioned above,
 one arrives at 
\bea\l{PQvdw-na}
\!\!P(T,n)=T r^{}_1\frac{n\left[1+\delta_{N}\right]}{1-\tilde{b}^{}_Nn}+
 T r^{}_2\frac{n\left[1-\delta_{\alpha}\right]}{1-\tilde{b}^{}_\alpha n}
-a^{}_1n^2,
\eea
where
\be\l{delta}
\delta_{i}=\frac{\epsilon^{}_{i}}{1-\tilde{b}_i n}, \quad
n^{}_N=r^{}_1\,n, \quad n_\alpha=r^{}_2\,n~,
\ee
and  $i=N,\alpha$,  $r^{}_1$ and $r^{}_2$ are given by
Eq.(\ref{tbest}).
Note that the  expression for the  pressure,
Eq.~(\ref{PQvdw-na}), in a case of $r^{}_2=0$ and $r^{}_1=1$ exactly the same as for
a pure
nuclear matter in Ref.~\cite{FMG19}.
 A new
     feature of the quantum
effects in the system of particle with the vdW interactions is the additional
factors $(1-\tilde{b}_{i}n)^{-1}$
in the quantum correction $\delta_i$,
i.e., the QS
effects becomes stronger due to the repulsive interactions between particles.

\begin{figure*}
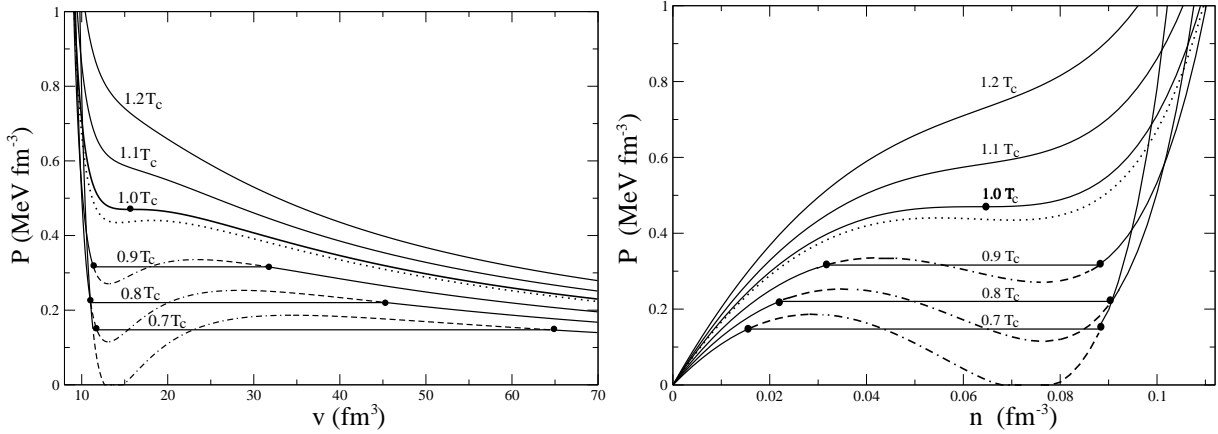

\begin{center}
\includegraphics[width=8.0cm,clip]{Fig3a_PvT.eps}
\includegraphics[width=8.0cm,clip]{Fig3b-PnT.eps}
\end{center}

\caption{
  Pressures $P$ as functions of the reduced volume $v$ (left) 
  and
particle number density
$n$ 
(right panel) at different temperatures $T$ (in units of the critical value $T_c$)
for the simplest case of the
symmetric nucleon matter. The critical point is shown by the
close circle found from the exact solution of 
    equations (\ref{CP-0}).
The dotted line shows the second order approximation
[Ref.~\cite{FMG19} and Eq.~(\ref{PQvdw-na}) ($r^{}_1=1, r^{}_2=0$)
      employing for nucleon matter].
The horizontal lines are plotted by using the Maxwell area
law in left and correspondingly in right panels.
The unstable and metastable parts of the isothermal lines are presented
by dashed and dash-dotted lines, respectively. 
    Other closed dots show schematically a binodal
boundary for the two phase coexistence curve in
the transition from two- to one-phase range \cite{FMG19}.
}
\label{fig3}
\end{figure*}
The vdW, both in its classical form (\ref{vdW}) and in its QvdW
extension (\ref{PQvdW}) and (\ref{PQvdw-na}), 
describes the first order liquid-gas
phase transition. As the value of $X_\alpha$, used in our derivations, is
very small, the approximate critical points in the considered approach will be
determined by the following equations:
\be\l{CP-0}
\left(\frac{\partial P(T,n)}{\partial n}\right)_T = 0~,~~~~
\left(\frac{\partial^2  P(T,n)}{\partial n^2}\right)_T=0~.
\ee
Using Eq.~(\ref{PQvdw-na}) in the first approximation in $\delta_i$,
one derives
from Eq.~(\ref{CP-0})
the
system of two equations
for the CP
parameters $n_c$ and $T_c$
at the same first order:
\bea 
&2na^{}_1 = \frac{T r^{}_1\,\left(1+2\delta^{}_N\right)}{\left(1-\tilde{b}^{}_1n\right)^2}
  +\frac{Tr^{}_2\,\left(1-2\delta_\alpha\right)}{\left(1-\tilde{b}^{}_2n\right)^2}~,\l{cp-1a}\\
    &a^{}_1 = \frac{T r^{}_1\tilde{b}^{}_1}{\left(1-\tilde{b}^{}_1n\right)^3}
\left[1+\delta^{}_N \frac{(1+2\tilde{b}^{}_1n)}{\tilde{b}^{}_1n}\right]\,\nonumber\\
    &+\frac{T r^{}_2\tilde{b}^{}_2}{\left(1-\tilde{b}^{}_2n\right)^3}
  \left[1-\delta_\alpha \frac{(1+2\tilde{b}^{}_2n)}{\tilde{b}^{}_2n}\right]~.
\l{cp-1b}  
\eea
Note that the   equations
    (\ref{cp-1a}) and  Eq.~(\ref{cp-1b}) for the CP
in 
    the case of $r^{}_2=0$ exactly the same as for a pure
nucleon 
matter in Ref.~\cite{FMG19}.
%

\vspace{0.3cm}
\begin{table}[pt]
\begin{center}
\begin{tabular}{|c|c|c|c|c|}
\hline
Critical points
&  vdW ($k_{\rm max}=0$) & 1 & 2 & QvdW  \\
\hline
$T_c$~[MeV] & ~29.2~& ~19.0~
&~19.67~& ~19.7\\
\hline
$n_c$~[fm$^{-3}$] &0.100 & 0.065
&~0.072~&~0.072\\
\hline
$P_c$~[MeV$\cdot$ fm$^{-3}$] & 1.09  & 0.48
&~0.52~&
~0.52
\\
\hline
\end{tabular}
\vspace{0.2cm}
\caption{{\small
Results for the CP parameters of the symmetric nuclear
matter ($g=4, m=938~\mbox{MeV}$); $k_{\rm max}=0, 1, 2$ is the order of
the QS
expansion where $k_{\rm max}=0 $ is the vdW approach (Eq.~(\ref{CP}))
(from Ref.~\cite{FMG19}).
}}
\label{table-1}
\end{center}
\end{table}
%

\vspace{0.3cm}
\begin{table*}[pt]
\begin{center}
\begin{tabular}{|c|c|c|c|c|c|}
\hline
Critical points  & Eq.(\ref{CP})
& $N$ 1st-order & $N$ QvdWM  & $N+\alpha$ 1st-order & $N+\alpha$ QvdWM\\
\hline
$T_c$~[MeV] & ~29.2~& ~19.0~
& ~19.7~&~19.4~&~19.9\\
\hline
$n_c$~[fm$^{-3}$] &0.100 & 0.065
& 0.072 &~0.072~&~0.073\\
\hline
$P_c$~[MeV$\cdot$ fm$^{-3}$] & 1.09  & 0.48
& 0.52 &~0.51~&~0.56\\
\hline
\end{tabular}
\vspace{0.2cm}
\caption{{\small
Results for the CP parameters of the vdWM (2nd column), the symmetric nuclear
matter ($N$) ($g=4, m=938~\mbox{MeV}$, 3rd and 4th columns) and
the mixed symmetric-nuclear
and $\alpha$-particle ($g=1, m=3737~\mbox{MeV}$) matter ($N+\alpha$,
5th and 6th columns).
}}
\label{table-2}
\end{center}
\end{table*}

For the CP parameters of the classical vdWM, 
which are found from Eq.~(\ref{CP-0})
for the equation (\ref{vdW}), one has 
\bea\l{CP}
  & T_c^{(0)}=\frac{8a}{27b}\cong 29.2~{\rm MeV}~, ~~~
  n_c^{(0)}=\frac{1}{3b}\cong 0.100~{\rm fm}^{-3}~,\nonumber \\
  & P_c^{(0)}=\frac{a}{27b^2}\cong 1.09~{\rm MeV}\cdot {\rm fm}^{-3}~.
  \eea
The numerical calculations within the full
QvdWM 
(\ref{PQvdW}), (\ref{PNal}), (\ref{nNal})
and 
(\ref{nN}) give (see also Refs.~\cite{marik,vova,FMG19})
\bea\l{CP-num}
  & T_c\cong 
  19.9~{\rm MeV}~, ~~~n_c\cong 
  0.0733~{\rm fm}^{-3}~,\nonumber \\
  & P_c\cong 
  0.562~{\rm MeV}\cdot {\rm fm}^{-3}~.
\eea
These our results
(\ref{CP-num})
appear to be essentially the same as
those  obtained in Ref.~\cite{vova}.

 A summary of the results for the CP parameters
 is presented in Tables \ref{table-1} (with Figs.~\ref{fig2} and \ref{fig3})
 and \ref{table-2}.  For symmetrical nuclear 
 matter ($X_\alpha=0$), Fig.~\ref{fig3} shows the isotherms of the pressure $P$
 as function of the reduced volume $v$ (left) and the particle number density $n$ (right panel)
 with the 
 first (and second) order corrections. For the same case,
a difference of the results for the  classical vdWM   
(\ref{CP}) and
QvdWM 
(\ref{CP-num})
demonstrates
a role of the effects of Fermi and Bose statistics at the CP of the symmetric
nuclear 
particle matter. The size of
these effects appears to
be rather significant for the case of impurity contributions 
$X_\alpha \cong 1$ of the
$\alpha$-particles into the nucleon matter.
On the other hand, it is remarkable that the  first order correction
(Table \ref{table-1})
reproduces these QS
effects
with  a high accuracy.  
The contribution of high order corrections in $\delta_i$, - second , third
and fourth order
is much
  smaller than the first-order correction                        
  that shows
  a fast convergence
  in $\delta_i$
  by 
  first-order terms. Therefore,  high-order 
  corrections due to the QS
  effects can be neglected
  for evaluations  of the critical points values.

  Table \ref{table-2} shows that for the case of the mixed $N-\alpha$
  system with $X_\alpha$, Eq.~(\ref{Xal}), even the first order
  corrections are in good agreement with exact numerical QvdW results
  (\ref{CP-num}), see Refs.~\cite{marik,vova,FMG19}. As seen from
  Table \ref{table-2}, the QS
  effects of the $\alpha$- particle impurity can be
  neglected because, first of all, of too small relative concentration
  $X_\alpha$ of this impurity, according to Eq.~(\ref{Xal}) as suggested in
  Ref.~\cite{vova}. By this reason, one can simplify our calculations of the
  particle number fluctuations in the next section \ref{sec-4}, taking a pure
  symmetric nuclear matter.

Many other examples
were recently considered in Ref.~\cite{VOV-17}.
All  models investigated in that paper have rather different 
 high-order virial-expansion coefficients.
However, if the parameters of these different models are fixed by a
requirement
to reproduce  properties of the 
ground state, the obtained values
of $T_c$
and $n_c$ appear to be quite similar. For example, different $T_c$ values
come to the narrow region
$T_c=18\pm 2$~MeV for the symmetric nuclear matter ($X_\alpha=0$).
The effects of Fermi statistics
leads to much stronger changes of the $T_c$ values: about 10~MeV in the
nucleon matter.

%
\begin{figure*}
\begin{center}
  \includegraphics[width=7.0cm,clip]{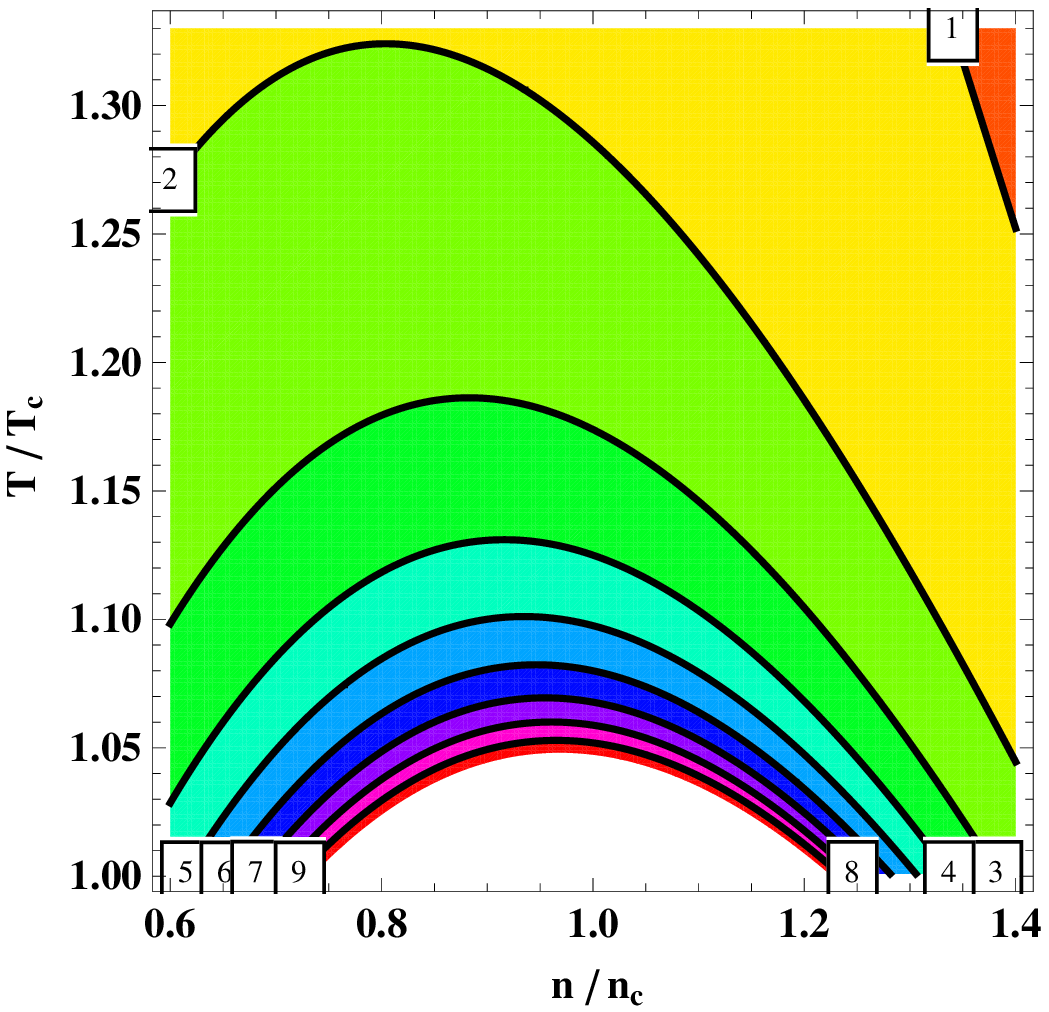}
  ~
  \includegraphics[width=7.0cm,clip]{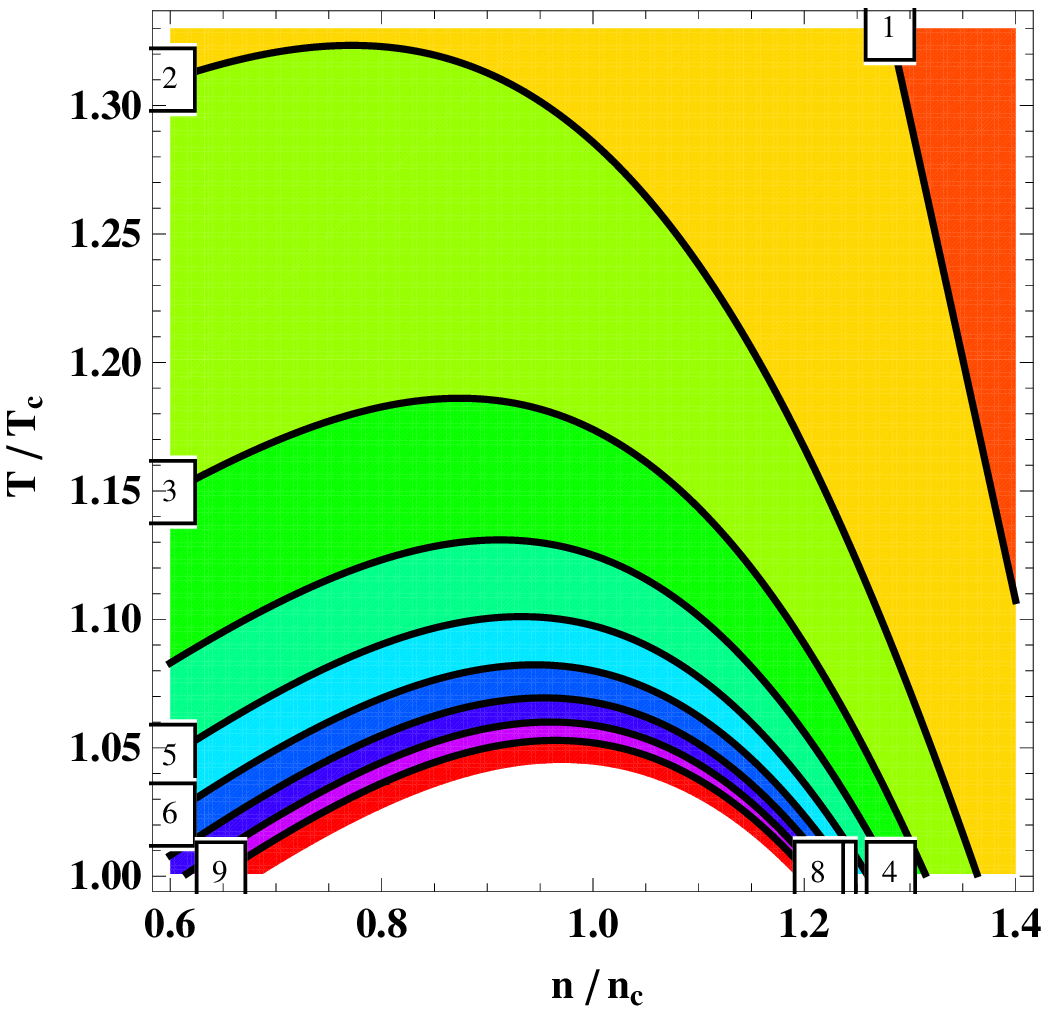}
  
  \includegraphics[width=7.1cm,clip]{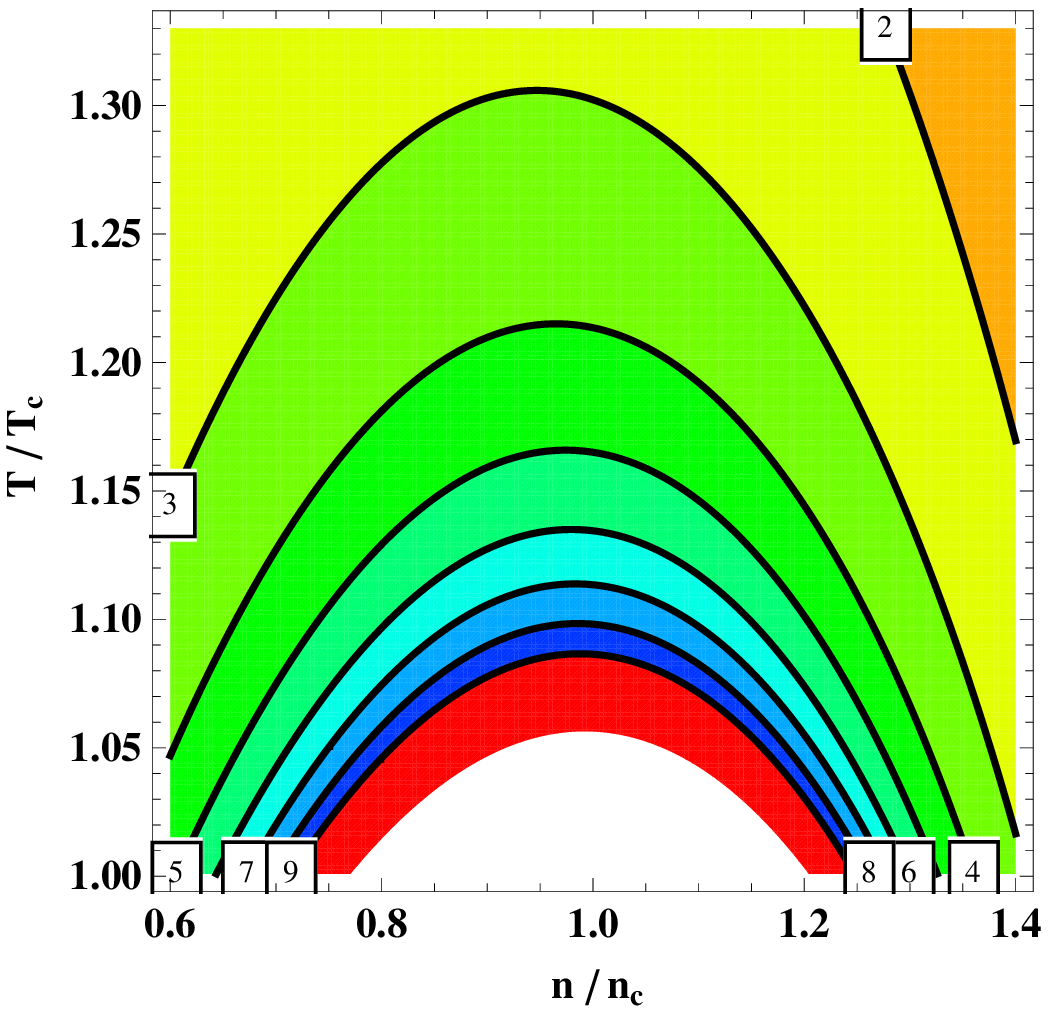}
  ~
    \includegraphics[width=7.1cm,clip]{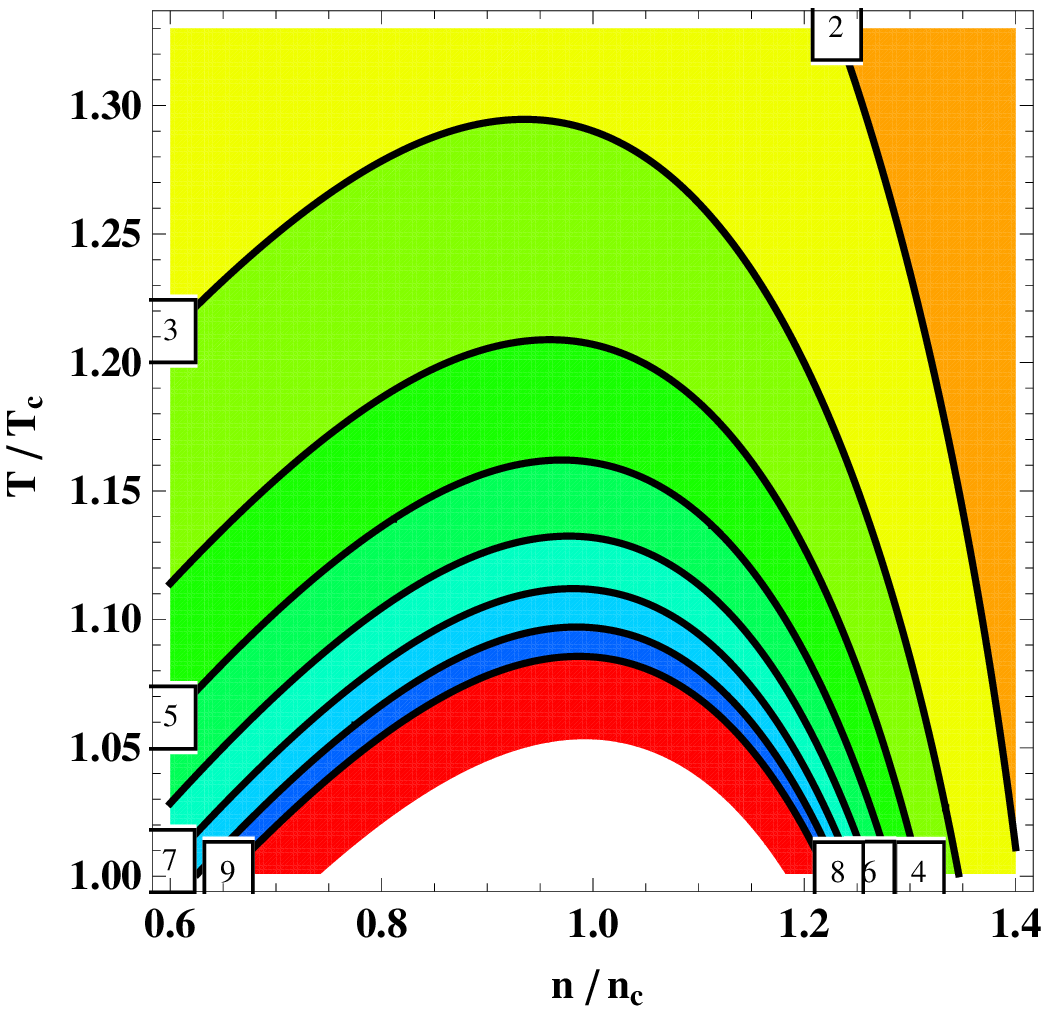}
\end{center}

\caption{
 Contour plots for the vdW [zeroth, upper ] and first (lower plots)
  orders in the QS
  expansion  over a small parameter $\delta$
  for the particle number fluctuations as functions of the
  density $n$ and temperature $ T $ 
  (in units of $n_c$ and $T_c $)
  with full in-compressibility $\mathcal{K}_T(n,T)$ (left) and the
  main derivative approximation (MDA) (right panels).
}
\label{fig4}
\end{figure*}

\section{ Particle number fluctuations}\label{sec-4}

From the Gibbs probability
distribution for a
gas of classical particles interacting through the repulsive and
attractive forces at large temperatures $T$ and small enough particle number
density average $\langle n\rangle$, one can use the vdW equation
of state (\ref{vdW})
[see sec. \ref{sec-2} and Ref.~\cite{LLv5})].
In what follows, to simplify notations,
we will omit angle brackets for statistical averages if it does not
lead to misunderstanding.
Following, e.g., Ref.~\cite{ZM02},
the fluctuations of the particle number $\omega$ as the dispersion of the
GCE Gibbs distribution function integrated over the excitation energy
can be expressed in terms of the derivative of particle number-density average.

\subsection{Fluctuations and susceptibility in the GCE}
\l{subsec-4a}

For calculations of classical fluctuations of the particle
numbers, $\omega$, within the grand canonical ensemble (GCE)
one can start with the particle number average \cite{ZM02,LLv5}
\be\l{avpartnumb}
\langle N \rangle=\int \rho^{}_{N}({\bf q},{\bf p}; \mu, T, V) \d \Gamma~,
\ee
where $\rho^{}_{N}({\bf q},{\bf p}; \mu, T, V)$ is the GCE distribution function
of the phase space variables ${\bf q},{\bf p}$, $\d \Gamma=
\d {\bf q}\d {\bf p}$ (normalized as usually for a classical system),
 $\mu$ is the chemical potential, $T$ the
temperature,  and $V$ the volume of the classical system. The
Gibbs probability distribution can be written as  
\be\l{distfun}
\rho^{}_{N}({\bf q},{\bf p}; \mu, T, V)=\mathcal{Z}^{-1}
\exp\left[-\left(H_N({\bf q},{\bf p}) -\mu N\right)/T\right]~.
\ee
Here $H_N({\bf q},{\bf p})$ is the classical Hamiltonian, $\mathcal{Z}$ the
normalization factor which is the partition function,
\be\l{partfun}
\mathcal{Z}=
\sum_N \int \d\Gamma\exp\left[-\left(H_N({\bf q},{\bf p}) -\mu N\right)/T\right] ~.
\ee
Taking variations of both sides of Eq.~(\ref{avpartnumb}) over $\mu$ with the
help of
Eqs.~(\ref{distfun}) and (\ref{partfun}) and changing the order of the integral
over the phase space $\Gamma$ and derivative over the chemical potential $\mu$,
at first order variations, i.e., the  second order in fluctuations one obtains
(see Refs.~\cite{LLv5,ZM02})
\begin{equation}
  \omega (n,T) =
         \frac{\langle \left (\Delta N \right)^2\rangle}
         {\langle N \rangle}=
         T\frac{(\delta n/\delta \mu)^{}_T}{n}~,
 \label{FL-sus}
\end{equation}
where $\Delta N=N-\langle N \rangle$ is the fluctuation of $N$ around its average  $\langle N \rangle$, $n=n(\mu,T)$ is the particle number-density average in the
GCE.  In Eq.~(\ref{FL-1sus}), the 
variational derivative,
\be\l{chi}
\chi=(\delta n/\delta \mu)^{}_T~,
\ee
is the isothermal susceptibility. For the linear (first order) variations,
\be\l{chi1}
\chi^{}_1=(\partial n/\partial \mu)^{}_T~,
\ee
one has explicitly,
\begin{equation}
  \omega^{}_1 (n,T)\equiv S_2=
         \frac{\langle N^2\rangle-\langle N \rangle^2}
         {\langle N \rangle}=
         T\frac{(\partial n/\partial \mu)^{}_T}{n}~.
 \label{FL-1sus}
\end{equation}
The particle number density
$n(\mu,T)$, entropy density $s(\mu,T))$, and energy density
$\varepsilon(T,\mu)$  in the GCE are given
by Eq.~(\ref{term}).

 Let us consider variations of the relationship (\ref{avpartnumb}) over the
 chemical potential $\mu$ taking into account high order variations,
 for instance,
 second-order ones. For simplicity, we shall still take these variations
 at constant
 temperature, i.e. consider non-linear (second-order)
 isothermal susceptibility. Eq.\ (\ref{FL-sus}) is correct for any order of the
 variational derivative (non-linear susceptibility, Eq.~(\ref{chi})) but now we
 can specify it at the 2nd order.
 Taking immediately the  variations over $\mu$ up to the second order
 at $T=const$ in Eq.~(\ref{avpartnumb}), one obtains
 the next (2nd) order corrections to Eqs.\ (\ref{FL-1sus})
 and (\ref{chi1}), which were considered at first order.
 These corrections are proportional
 to the so called  kurtosis, defined in Ref.~\cite{marik}
 in a slightly different way. 
 Fluctuations 
accounting for the third cumulant moment of the Gibbs distribution, take the form:
 \be\l{2ndordervar}
 \frac{T}{\langle N \rangle}\delta_2^{}\langle N \rangle=
  S_2~(\delta \mu)^{1} 
 +\frac{1}{2T}  S_3~(\delta \mu)^2 +\ldots~,
 \ee
 where $S_3$ is the kurtosis 
 which can be normalized by the $\langle N\rangle$ as
 $S_2$, Eq.\ (\ref{FL-1sus}),
 \be\l{kurtosis}
S_3=\frac{\langle N^3\rangle -
 \langle N\rangle^3 }{\langle N \rangle}~.
 \ee
Similarly, one can obtain the 4th order moment (or 4th-order cumulant moment) of the
 Gibbs distribution, i.e.,
from the third order
 variations of the average  (\ref{avpartnumb}) over chemical potential
 $\mu$, 
 and so on.
 This allows us to go beyond the restrictions of the 2nd order cumulant moment
 fluctuations $\omega^{}_1$, shown explicitly in Eq.\ (\ref{FL-1sus}), i.e.
 beyond the first variational derivative for the
 susceptibility $\chi$,
 -- linear susceptibility $\chi^{}_1$, Eq.~(\ref{chi1}).

 The expression (\ref{FL-sus})
 for the fluctuation $\omega$ of the particle number is more general though
 it is still singular exactly at the CP where the linear susceptibility
 $\chi^{}_1$ (\ref{chi1}) is $\infty$ in the sum (\ref{2ndordervar}).

The 
integral traces of cumulants as given by 
\be\l{cum}
C_N=\int \d \q \d \p~
\exp\left\{-\beta
  \left[H_N({\bf q},{\bf p}) -\mu N\right]\right\}~,
  \ee
 can be calculated  by the saddle point method (SDM).
We may try to introduce the entropy
\be\l{S}
S=-\sum^{}_N P_N\log P_N~,
\ee
where $\mathcal{P}_N=\rho^{}_N(\q,\p,\mu,T,V)$ is the probability
distribution (\ref{distfun}) 
\be\l{PF}
\mathcal{P}_N=
\exp\left\{-\beta \left[H_N({\bf q},{\bf p})-E\right]-\mu N\right\}~,
\ee

Writing the SPM condition
$\delta S=0$, i.e.,
\be\l{SPC}
\left(\frac{\partial S}{\partial \q}\right)^\ast=0, \qquad
\left(\frac{\partial S}{\partial \p}\right)^\ast=0,
\ee
one obtains the classical trajectories $\q^\ast(t), \p^\ast(t)$ from
these Hamiltotian equations (\ref{SPC}). We have also to specify the mean field in
the Hamilton function,
$H_N({\bf q},{\bf p})=\sum_\kappa\left[\p^2_\kappa/2m + V_\kappa(\q_\kappa)\right])$.
If we are
far from the bifurcations
(CPs), one can use the standard SPM, and the non-zero 2nd order terms
of the entropy expansion.
Such derivations lead to 
the results of the standard thermodynamics but near the CP.
Near the
bifurcation, where the second order terms of the entropy expansion is zero, we
may employ the improved SPM (ISPM) \cite{mafm,MY11}
transforming  the ISPM for the action
phase integral of the POT \cite{gutzpr,gutzbook,strumag,BB03,MY11} to the real
exponent argument -- the entropy $S$,
Eq.\ (\ref{S}). The simplest ISPM is the second - order expansion of the
entropy (\ref{S}), but with finite integration limits.
Now, one can take the path integral 
analytically
in terms of the $\mbox{erf}$ functions of the real argument.
Here is the place where we can apply for the catastrophe theory of Fedoryuk
\cite{fed:jvmp,maf} by expanding the entropy
to the third order terms. In this way, we arrive at the Airy-kind integrals
with the finite contributions of the two SPM points  which turn into one
bifurcation point at the limit to the CP.
Note that in order to remove singularity of the fluctuations $\omega$ near at the critical point
with generalization to the QS description, one can  calculate $\omega$
through the moments of the statistical level density $\rho(E,A) $ \cite{MS20}, also with
using the ISPM.

Thus, for the first simplest classical dynamic case, the integral for
the fluctuation (\ref{FL-sus}) can be presented
as the Feynman path integral over the formal
trajectories $\q(t)$ and $\p(t)$ with the SDM condition (\ref{SPC})
for the main contributions from the classical trajectories at large excitation
energy with the system temperature,
$T=\left(\partial S/\partial E\right)^\ast$, where the standard thermodynamics
  but with critical points is working well.

  In the next sections, we
  will study more a popular formula (see Appendix A,
  Ref.~\cite{LLv5,IA71,BR75})
  used for calculations of the fluctuations $\omega$,
  which is  expressed in terms of the isothermal in-compressibility
  $\mathcal{K}_T$,
  and compare the results obtained by different approximations.
  Our purpose of the next sections is
  to find the ranges of good agreement between the approximate expansion near
  the critical point and accurate analytical result for the vdWM 
  to check a validness of both  expressions through the non-linear
  susceptibility $\chi$ and non-linear in-compressibility
  $\mathcal{K}_T$.
%
\begin{figure*}
\begin{center}
   \includegraphics[width=10.0cm,clip]{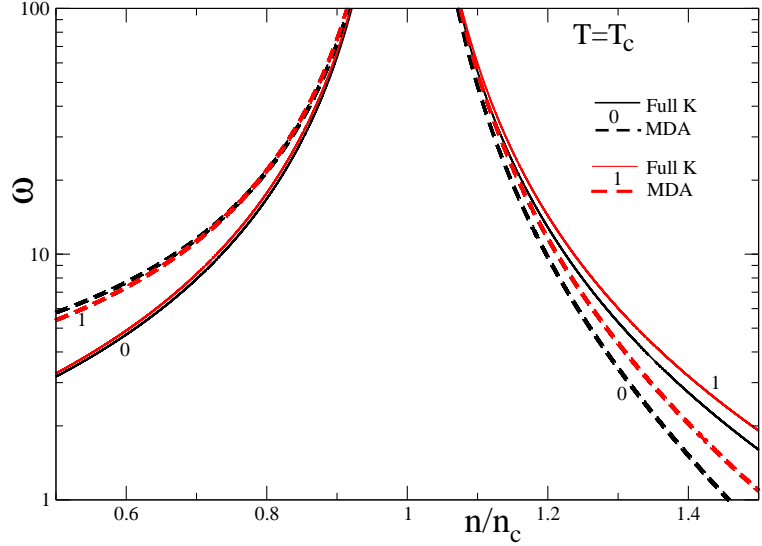}
\end{center}

\caption{
  Fluctuations of the particle numbers $\omega$
  for nucleon system as
  function of the particle
  number density $n$ 
  (in units of $n_c$) at
  the critical value of the temperature  $T=T_{c}$  
  with zeroth (vdW), $k_{\rm max}=0$, and first corrections 
  of the QS expansion, $k_{\rm max}=1$.
   Solids show Eq.~(\ref{FL-sus}) for the full
  (without expansion near the CP) fluctuations $\omega$
    and dashed curves
  present the tested main-derivative approximation (MDA),
  Eq.~(\ref{FL-press})
  with (\ref{incompexp}).
  }
\label{fig5}
\end{figure*}

\subsection{Fluctuations and in-compressibility}
\l{subsec-4b}

For the relative fluctuations  of the
particle numbers $\omega$,
one  has
\cite{TR38,RJ58,LLv5,TK66,IA71,BR75,AC90,ZM02}
\begin{equation}
  \omega (n,T) = \frac{T}{\mathcal{K}_T}~,
 \label{FL-press}
\end{equation}
where $\mathcal{K}_T $ is the isothermal in-compressibility,
\be\l{KT}
\mathcal{K}_T = \left(\frac{\delta P }{\delta n }\right)_T~,
\ee
and
$P$ is given by equation of state which is given in the one-component
QvdW (symmetric nucleon matter) by
Eq.~(\ref{PQvdw-na}) (with $r^{}_1 = 1$, $r^{}_2 = 0$,
 $\tilde{b} = b^{}_{NN}$).
The in-compressibility $\mathcal{K}_T$, Eq.~(\ref{KT}),
in Eq.~(\ref{FL-press}) as function of the density $n$ and temperature $T$,
can be expanded in power
series near the critical point $n_c, T_c$ over both variables $n$ and $T$
but taking derivatives at the current point $n,T$,
\bea\l{incompexpfull}
&\mathcal{K}_T
=\left(\frac{\partial P}{\partial n}\right)_T +
\left(\frac{\partial^2 P}{\partial n^2}\right)_T (n-n_c)\nonumber\\
&+
  \frac{\partial^2 P}{\partial n\partial T}(T-T_c) +
  \frac12\left(\frac{\partial^3 P}{\partial n^3}\right)_T
  \left(n-n_c\right)^2+\ldots ~.
\eea
Using approximately the definition (\ref{CP-0}) valid
at the critical
point\footnote{The CP is assumed to be of the simplest second order,
  in contrast to a high order
CP when high order derivatives become also zero.}, and assuming that the linear in
temperature  and 
quadratic in density variations are dominating
above other high order variations, one can define the main derivative
approximation (MDA):
\be\l{incompexp}
\mathcal{K}^{\rm MDA}_T\approx
  \frac{\partial^2 P}{\partial n\partial T}(T-T_c) +
  \frac12\left(\frac{\partial^3 P}{\partial n^3}\right)_T
  \left(n-n_c\right)^2~.
\ee
We may compare
their
contributions  $\mathcal{K}^{\rm MDA}_T $ into the full expansion
(\ref{incompexpfull}) 
with the definition $\mathcal{K}_T(n,T)$, Eq.~(\ref{KT})
in terms of the first order derivative in the expansion (\ref{incompexpfull})
as function of $n,T$,
\be\l{incompexp1}
\mathcal{K}_T \approx \mathcal{K}^{(1)}_T
=\left(\frac{\partial P}{\partial n}\right)_T~.
\ee
Notice that Eq.~(\ref{FL-press}) can be derived from Eq.~(\ref{FL-sus})
by using linear variations for the chemical potential $\mu$ as function
of the particle number density $n$ (see, e.g., Appendix \ref{sec-der}).

Approximating 
Eq.~(\ref{KT}) for  in-compressibility $\mathcal{K}_T(n,T)$ by
the first derivative of the pressure, Eq.~(\ref{incompexp1}),
at the first order in a small
quantum-statistics parameter $\delta$,  Eq.~(\ref{delta}) ($r^{}_1 =
1$, $r^{}_2 = 0$, $\tilde{b} = b^{}_{NN}=b$, Eq.~(\ref{bij})),
one obtains
\begin{equation}
    \omega (n,T) =\frac{T}{T [1+2\delta]/(1-nb)^2-2na}~.
 \label{omega2}
\end{equation}

Studying now
a behavior of  $\omega (T,n)$, Eq.~(\ref{omega2}), near the critical point
$n_c,T_c$, see 3rd column in Table \ref{table-1}, within the QvdW model,
we will use now the expansion of these fluctuations in powers of the distance
from the CP $n_c,T_c$ taking derivatives at the CP.
  With the help of the new variables,
\be\l{taunu}
  \tau \equiv ~T/T_c-1~, ~~~~
  \nu \equiv ~n/n_c 
  -1~,
  \ee
    one can fix first
    $n=n_c$ and
        find the behavior  of $\omega(n,T)$ as function of temperature $T$
    near the critical point.
For this purpose, it is convenient to present
$\delta (n,T)$ in the following form:
 \begin{equation}
        \delta (n,T) ~\approx \delta \left[(1+\nu) n_c,(1+\tau) T_c\right].
    \label{deltataunu}
  \end{equation}
 We will find now the limit of this expression at $\nu=0$ and very
 small $\tau$ and,
     then, at $\tau=0$ and very small $\nu$. In the first case, $\nu=0$, 
 one can approximate Eq.~(\ref{deltataunu}) by
  \begin{equation}
    \delta (n_c,(1+\tau)T_c) \approx
    \frac{\hbar^3 \pi^{3/2}\tilde{n}}{2g~(mT_c)^{3/ 2}\left(1-\tilde{n}\right)}
    \left(1- \frac{3}{2}\tau \right)~.   
 \label{deltatau}
  \end{equation}
Using also Eq.~(\ref{omega2}), one finds
\bea\label{omegaproxtau}
 &  \omega(n_c,(1+\tau) T_c) \approx
  \frac{\left(1-\tilde{n}\right)^2}{1-
    \delta^{}_0}~~\tau^{-1}~\nonumber\\
 & =~\frac{T_c n_c}{P_c } ~ G_\tau~ \tau^{-1}~,
    ~\nu=0~,
   \eea
where  $\tilde{n}=b n_c$,  
$\delta^{}_0=
\delta(T_c,n_c)$, and 
  \begin{equation}
G_\tau \approx \frac{P_c}
    {T_c n_c}~\frac{\left(1-\tilde{n}\right)^2}{1-
        \delta^{}_0}~.    
 \label{Gtau}
\end{equation}
Taking $b$ from Eq.~(\ref{bij}) , one
finally obtains 
$G_\tau \approx 0.29$, that is only slightly different from the
value $G_\tau \approx 0.26$
of Ref.~\cite{roma}.
For the case of the classical vdWM, one respectively 
arrives at $G_\tau= 1/6$.

Similarly, 
using Eq.~(\ref{deltatau}), for the fluctuations $\omega(n,T) $, Eq.~(\ref{omega2}),
at the constant $T=T_c$
one  finds 
\bea\l{deltanu}
 &\delta ((1+\nu)n_c,T_c) ~\approx
    ~\frac{\hbar^3 \pi^{3/2}\,\tilde{n}}{2g~(mT_c)^{3/ 2}\left(1-\tilde{n}\right)}~\nonumber\\ 
  &\times \left(1+ \frac{\nu}{1-\tilde{n}}+
        \frac{\nu^2~\tilde{n}}{(1-\tilde{n})^2} \!\right). 
 \eea
Finally, for the fluctuations $\omega$, Eq.~(\ref{omega2}),  one arrives at
\begin{equation}\l{omegaproxnu}
  \omega ((1+\nu)n_c,T_c) \approx ~\frac{T_c n_c}{P_c }~ G_\nu
    ~\nu^{-2}~,~~~\tau=0~,
\end{equation}
where
\be\l{Gnu}
G_\nu\approx ~\frac{P_c}{T_c n_c }~~
 \frac{\left(1-\tilde{n}\right)^4}{3\tilde{n}~[
      2\delta^{}_0\left(1+\tilde{n}\right)
            +\tilde{n}]} \approx 0.33~. 
 \ee
 For the case of the classical vdWM ($\delta^{}_0=0$), from Eq.~(\ref{Gnu}) one obtains
 $G_\nu= 2/9$ which is the same as that of Ref.~\cite{marik}.
 As the fluctuation $\omega(n,T)$,  Eqs.~(\ref{FL-press}) or (\ref{omega2}),
 is function of the two variables $n$ and $T$,
 one needs to introduce the two-dimensional critical index,
 with the first 
 component 
 being along the $n$ and second one 
 along the $T$ axis. Another characteristics of the
critical point ($n_c,T_c$) in the $n-T$ plane is the two-dimensional fluctuation slope
 coefficient 
$\{G_\tau,G_\nu\}\approx\{0.29,0.33\}$. 
 Notice that the temperature, Eq.~(\ref{omegaproxtau}),
and the density, Eq.~(\ref{omegaproxnu}), dependence near the CP can be seen
also from Eq.~(\ref{incompexp}) of the MDA.

\subsection{Discussion of the results}
\l{sec-5}

Fig.~\ref{fig4} shows the particle number fluctuations
$\omega(n,T)$ in units of the critical values $n_c$ and $T_c$
for symmetric nuclear matter at the zeroth
[vdW, upper] and first (lower panels) order  in the quantum statistics
expansion.
Left  and right contour plots of Fig.~\ref{fig4} present the
calculations using respectively the
standard in-compressibility $\mathcal{K}^{(1)}_T(n,T)$, Eq.~(\ref{incompexp1}),
and its MDA, Eq.~(\ref{incompexp}).
For the MDA
calculations we assume
the dominance of the derivative contributions of Eq.~(\ref{incompexpfull})
above high order variations in the in-compressibility,
see Eq.~(\ref{incompexp}), and neglect first- and second-derivative terms
by using approximately Eq.~(\ref{CP-0}). As seen from Fig.~\ref{fig4}
(cf. lower with upper plots),
the quantum statistics effects is significant for the fluctuations $\omega$
even after exclusion of
a large shift of the critical point by choosing the scaling units
to a lower critical values due to the quantum statistics effect, in agreement
with the accurate
numerical result [Eq.(\ref{CP-num})] (see also Ref.~\cite{vova}).
Contour plots for fluctuations $\omega$ at a
few next high orders
(e.g., $k_{\rm max}=2-4$)
are almost the same as for the first order and, therefore, is not shown in
Fig.~\ref{fig4}.
It is clearly seen that a convergence of the MDA fluctuations
$\omega$, Eqs.~(\ref{FL-press}) and (\ref{incompexp}), with those calculated
through the 
the equation~(\ref{incompexp1}) for in-compressibility $\mathcal{K}_T$
takes place,
except for small white ranges near the CP.
(see Fig. \ref{fig4}). 

Fig.~\ref{fig5} presents more details in the comparison 
between fluctuations 
(\ref{FL-press}) with the in-compressibility
$\mathcal{K}_T$, Eq.~(\ref{incompexp1}),
using the pressure (\ref{PQvdw-na}) ($r^{}_1=1, r^{}_2=0$) for nucleons
at the zeroth
and 
first 
orders of the QS expansion 
and their MDA calculations by Eq.~(\ref{incompexp}) for $\mathcal{K}$, 
at $T=T_c$. 
The
  derivatives of the MDA are 
  calculated analytically at the ($n,T$) point on a small but
  finite distance from the
 critical point ($n_c,T_c$)
by assuming that the third derivative term over the density $n$ is leading 
 at  the temperature $T=T_c$ (Eq.~(\ref{incompexp}))
 for variations of the pressure of equation of state,
which is determined by  Eq.~(\ref{PQvdw-na}) for symmetric nuclear matter.
As shown in 
Fig.~\ref{fig5}, 
the fluctuations (\ref{FL-press})
calculated through the 
in-compressibility, Eq.~(\ref{incompexp1}) (solids),
and the MDA, Eq.~(\ref{incompexp})(dashed lines), within a
given order $k_{\rm max}=0$ or $1$ of the QS expansion
shows a huge bump in 
the
density 
dependence, 
largely in agreement with the
approximate
simple analytical asymptotic expression (\ref{omegaproxnu}), and  more accurate
analytical formula, Eq.~(\ref{omega2}), also with numerical
calculations.

Concerning 
the MDA,  
one finds a good agreement with the expression 
(\ref{incompexp1}) for the vdW
($k_{\rm max}=0$), and first
(''1'') 
approximations in the QS expansion over
$\delta$ near the critical point up to a small distance from the CP. 
As this
distance decreases, 
one can see a divergence
    of the MDA as for the vdW approximation. For larger distances from the CP
    in the range $n \simg 1.5$ and $n \siml 0.5$, the discrepancy between these
    approximations are due to the fact that the MDA, Eq.~(\ref{incompexp}),
    becomes worse because these MDA high-order derivative contributions
    are not already dominating above the first two derivatives terms of
    Eq.~(\ref{incompexpfull}).
    Note that for the calculations of fluctuations $\omega$
    with QS corrections, it was convenient to use the expression
    (\ref{FL-1sus}) for the fluctuation $\omega$ in terms of the
    susceptibility using the fugacity variable $z$
    instead of the particle density variable $n$. Similarly, one can consider
the fluctuations $\omega(T,n)$ at $n=n_c$ with analogous properties.

A validity of the expressions (\ref{FL-press})
and (\ref{FL-1sus}) for fluctuations $\omega$ and their MDAs
can be evaluated from these calculations (Figs.~\ref{fig4} 
and \ref{fig5}) by the ranges where one finds good agreement
between the approximations (\ref{incompexp}) and (\ref{incompexp1})
for the in-compressibility $\mathcal{K}(n,T)$.
Their rough relative estimates, about 1.5\%, are found
approximately for both the cases,
$T=T_c$ (Fig.~\ref{fig5}) in the dependence on density
$n$ and in dependence on temperature at 
$n=n_c$
(see also Fig.~\ref{fig4}). 
The fluctuation based on the MDA Eq.~(\ref{incompexp}), converges to
that with the standard Eq.~(\ref{incompexp1})
for the in-compressibility
of the expression (\ref{FL-1sus}) (or Eq.~(\ref{FL-press}))
on much smaller relative distances, 0.05\% for $T=T_c $ and
from -0.02\% to 0.005\% for $n=n_c$,
with respect to the
critical values of Table \ref{table-1}.

Notice 
that it is difficult (impossible) to realize practically
the conditions for the
application of the MDA in the limit to the CP, 
in particular, if we introduce
the restrictions  $T=T_c$ or $n=n_c$. In the way to the CP, one has to stop at small but finite
distance from the CP when a huge bump appear : the MDA variations fail because it becomes smaller
or of the order of next derivatives contributions in expansion
(\ref{incompexpfull}) of the in-compressibility $\mathcal{K}_T$ in the
denominator of the fluctuations $\omega$, Eq.~(\ref{FL-press}), see
Refs.~\cite{TR38,RJ58,TK66}.
The derivations of Eqs.~(\ref{FL-sus}) and (\ref{FL-press})
become invalid on enough small but finite
distances from
the CP because, probably, we use the mean field approach (in particular, the vdWM) as the basis of the QS
perturbation expansion. As shown in Ref.~\cite{LLv5},
in this case the correlation length
of the correlation function, or
the two-body amplitude of scattering in the quasi-particle Landau theory
\cite{LP04}, infinitely 
diverges
by increasing relatively, in the considered limit to the CP, with respect
to the mean distance between
particles. In this case, the
arguments of validness
for the derivations of Eqs.~(\ref{FL-sus}) and (\ref{FL-press})
for the fluctuations $\omega$ through the derivatives of
the thermodynamic averages (pressure or particle number density)
contradict \cite{TR38,RJ58}
with the 
background of the
statistical physics for which we should have an opposite tendency such that
the considered relative fluctuations must be 
small; see, e.g.,
Refs.~\cite{TR38,RJ58,LLv5,ZM02}.

\section{Summary}\label{sec-6}

The QvdWM equation of state has been derived analytically and used to study
the quantum statistics effects
in a vicinity of the critical point of two-component system of
nucleon and $\alpha$-particle matter. 
The 
expressions for the pressure were obtained by using the quantum statistics expansion  ,
over the  small
parameters $\delta_i$ ($i=\{N,\alpha\}$) near the vdW approach. 
A simple and explicit dependence on the system parameters,
such as the
    particle mass $m_i$ and degeneracy factor $g_i$,
is demonstrated at the
first order of this expansion.
Such a dependence is absent within the classical vdWM. 
The quantum corrections to the CP parameters of the symmetric-nuclear
and $\alpha$-particle matter
appear to be quite significant.
For example, the value of $T_c^{(0)}=29.2$~MeV in the classical vdW model
{\it decreases} dramatically
to the value $T_c^{(1)}=19.4$~MeV. On the other hand,
this approximate analytical  result within the first-order
quantum correction is already  close to the accurate
numerical value of $T_c=19.9$~MeV
obtained by the numerical calculations within the full QvdWM. 
The trend of the critical-value changes because of inclusion of the
$\alpha$ particles into the nucleon system
occurs in the correct direction, namely the CPs are  somewhat increased
in the critical point as compared to those for pure nucleon system, and these
analytical results are
in good agreement with more exact numerical calculations.

The particle number fluctuations for symmetric nucleon matter
have been derived within the same analytical QvdW
approach near the critical point. Their behavior near the critical point
in standard calculations through the in-compressibility is in good agreement
with more exact numerical calculations. Main features, as a huge
bump near the CP, for the same QvdWM equation of state 
was found as similar to the approximate analytical and full numerical results
obtained with and without using the
expansion of the in-compressibility near the CP at zero (vdW) and first order
over a small parameter of quantum statistics. The convergence of the
main derivative approximation
for the isothermal in-compressibility near the CP by accounting for
contributions of the 
mixed second density-temperature and third density derivative terms
to the corresponding full expansion of the in-compressibility
was studied and the rough estimates for ranges of
validness
of these QvdW approximations was obtained.

As perspectives,
    we will study the fluctuations near the critical point
by using the improved saddle point method similarly as applied for the
oscillating components of the single-particle density of
states within the semiclassical periodic orbit theory of
critical points (bifurcations) \cite{mafm,maf,MY11}
and in terms of the moments of the statistical level density.
Note also that our consideration made for
the QvdWM 
can be straightforwardly
extended to other types of inter-particle interactions.

\begin{acknowledgments}
We thank A.I.~ Sanzhur for many fruitful discussions and suggestions, as well 
D.V.~Anchishkin, M.I.~Gorenstein,
A.~Motornenko, R.V.~Poberezhnyuk, and V.~Vovchenko for many useful discussions.
The work of S.N.F. and A.G.M. on the project
``Nuclear collective dynamics for high temperatures and neutron-proton asymmetries'' was
supported in part by the Program ``Fundamental researches in high energy physics
and nuclear physics (international collaboration)''
at the Department of Nuclear Physics and Energy of the National
Academy of Sciences of Ukraine.  S.N.F., A.G.M. and U.V.G. thank
the support in part by the budget program ``Support for the
development of priority   areas of scientific reseraches'', the project of the Academy
of Sciences of Ukraine, Code 6541230.
\end{acknowledgments}

\appendix

\section{Derivations of the classical particle-number fluctuactions}
\l{sec-der}

Within the canonical ensemble (CE), one can use the free energy $F(V,T)
$ as a characteristic thermodynamic function of the volume $V$ and
temperature $T$ for a fixed particle number
$N$. Assuming the thermodynamic limit condition for our infinite
system, one can express $F$ in terms of that per particle \cite{LLv5},
\be\l{f}
F(V,T)=Nf(\tilde{v},T)~,
\ee
where
\be\l{vtilde}
\tilde{v}=\frac{1}{n},~~~ n=N/V~.
\ee
For the pressure $P$ and chemical potential $\mu$, one has
\be\l{P}
P=-\left(\frac{\partial F}{\partial V}\right)^{}_T=
-\left(\frac{\partial f}{\partial \tilde{v}}\right)^{}_T~,
\ee
and
\be\l{mu}
\mu=\left(\frac{\partial F}{\partial N}\right)^{}_T=f - \frac{1}{n}
\left(\frac{\partial f}{\partial \tilde{v}}\right)^{}_T~,
\ee
where the volume per particle $\tilde{v}$ is given by Eq.~(\ref{vtilde}).

Taking the first variation of Eq.~(\ref{mu}) over particle number
density $n$ through
the relationship (\ref{vtilde}), one obtains
\be\l{dmudn}
\delta \mu=
\frac{1}{n^3}\left(\frac{\partial^2 f}{\partial \tilde{v}^2}\right)^{}_T~
\delta n~~.
\ee
Therefore, one finds
\be\l{dndmu}
\left(\frac{\partial n}{\partial \mu}\right)^{}_T=
\frac{n^3}{\left(\partial^2 f/\partial \tilde{v}^2\right)^{}_T}~.
\ee
According to Eq.~(\ref{FL-1sus}) and Eqs.~(\ref{dndmu}), (\ref{P}) and
(\ref{vtilde}),
one arrives at Eq.~(\ref{FL-press}).

Note that the same result can be obtained much shortly by using the
Jacobian transformations within the GCE \cite{LLv5},
\be\l{Jactrans1}
\left(\frac{\partial n}{\partial \mu}\right)^{}_T=\frac{D(n,T)}{D(\mu,T)}=
\frac{1}{D(\mu,T)/D(n,T)}~
\ee
and
\be\l{nGCE}
n=\left(\frac{\partial P}{\partial \mu}\right)^{}_T =~
\frac{D(P,T)}{D(\mu,T)}~,
\ee
see Eq.(\ref{term}). Therefore, substituting these equations
(\ref{Jactrans1}) and (\ref{nGCE})  into Eq.~(\ref{FL-1sus})
for the particle number fluctuations $\omega$,
one can do cancellation in ratios of the denominator 
by using the Jacobian properties. Finally, 
one obtains Eq.~(\ref{FL-press}).

Note that these derivations based on the first derivative
transformations fail near the
critical point because of the divergence of fluctuations due to zeros
in the denominators and, therefore, strictly speaking, cannot be used in
enough a small
vicinity of the critical point, see Eq.~(\ref{CP-0}), in contrast to
the fluctuation formula as the Gibbs distribution dispersion
(sec.~\ref{sec-3}) and Eq.~(\ref{FL-sus}) in terms of the susceptibility
(\ref{chi}).

As stated in the paper, our analysis can be applied 
beyond the vdW approach. 
In fact, similar estimates
of the quantum statistic effects can be straightforwardly
done also 
for the mean-field
models.  Concerning these models see, e.g.,
\cite{AV-15} and references therein.


\begin{thebibliography}{99}

\bibitem{nm-1} B. K. Jennings, S. Das Gupta, and N. Mobed, Phys. Rev. C {\bf 25}, 278 (1982).

\bibitem{nm-2} G. R\"opke, L. M\'unchow, and H. Schulz, Nucl. Phys. A {\bf 379}, 536 (1982).

\bibitem{nm-3} G. F\'ai and J. Randrup, Nucl. Phys. A {\bf 381}, 557 (1982).

\bibitem{nm-4} T. Biro, H. W. Barz, B. Lukacs, and J. Zimanyi, Phys. Rev. C {\bf 27}, 2695 (1983).

\bibitem{nm-5}  L. P. Csernai, H. St\"ocker, P. R. Subramanian, G. Buchwald,
  G. Graebner, A. Rosenhauer, J. A.
Maruhn, and W. Greiner, Phys. Rev. C {\bf 28}, 2001 (1983).

\bibitem{nm-6} L. P. Csernai and J. I. Kapusta, Phys. Rept. {\bf 131}, 223 (1986).

\bibitem{nm-7} B. D. Serot and J. D. Walecka, Adv. Nucl. Phys. {\bf 16}, 1 (1986).

\bibitem{nm-8} J. Zimanyi and S.A. Moszkowski, Phys. Rev. C {\bf 42}, 1416 (1990).

\bibitem{nm-9} R. Brockmann and R. Machleidt, Phys. Rev. C {\bf 42}, 1965 (1990).

\bibitem{nm-10} H. Mueller and B. D. Serot, Nucl. Phys. A {\bf 606}, 508 (1996).

\bibitem{nm-11} M. Bender, P. H. Heenen and P. G. Reinhard, Rev. Mod. Phys. {\bf 75}, 121 (2003).

\bibitem{ex-1} J. E. Finn et al., Phys. Rev. Lett. {\bf 49}, 1321 (1982).

\bibitem{ex-2} R. W. Minich et al., Phys. Lett. B {\bf 118}, 458 (1982).

\bibitem{ex-3}  A. S. Hirsch et al., Phys. Rev. C {\bf 29}, 508 (1984).

\bibitem{ex-4} J. Pochodzalla et al., Phys. Rev. Lett. {\bf 75}, 1040 (1995).

\bibitem{ex-5} J. B. Natowitz, K. Hagel, Y. Ma, M. Murray, L. Qin, R. Wada, and
  J. Wang, Phys. Rev. Lett.
{\bf 89}, 212701 (2002).

\bibitem{ex-6} V. A. Karnaukhov et al., Phys. Rev. C {\bf 67}, 011601 (2003).

\bibitem{vova}
  V. Vovchenko,  A. Motornenko, P. Alba, M.I. Gorenstein, L.M. Satarov,
  and H. Stoecker,
  Phys. Rev. C {\bf 96}, 045202 (2017).

  \bibitem{satarov} L.M.~Satarov, I.N.~Mishustin, A.~Motornenko, V.~Vovchenko,
    M.I.~Gorenstein, and H.~Stocker, Phys. Rev. C {\bf 99},
  024909 (2019).

\bibitem{roma1} R. V. Poberezhnyuk, V. Vovchenko, M. I. Gorenstein,
  and H. Stoecker
Phys. Rev. C 99, 024907 (2019).

   \bibitem{roma} R.V. Poberezhnyuk, V. Vovchenko, D.V. Anchishkin, M.I.
    Gorenstein,
    J. Mod. Phys. E, {\bf 26}, 1750061 (2017).  

\bibitem{marik} V.~Vovchenko, D.~Anchishkin, and M.~Gorenstein, Phys. Rev. C,
  {\bf 91}, 0.64314 (2015).

\bibitem{FMG19} S.N.~Fedotkin, A.G.~Magner, and M.I.~ Gorenstein,
  Phys. Rev. C, {\bf 100}, 054334 (2019).

\bibitem{AC90} M~~Anisimov and V.~Sychev,
  {\it Thermodynamics of critical state  for individual sustances},
  (Energoatomizdat, Moscow, 1990)(in Russian).
  
\bibitem{LLv5} L.D. Landau and E.M. Lifshitz, {\it Statistical Physics,
 Course of
 Theoretical Physics}, (Pergamon, Oxford, UK, 1975), Vol.~5.

 \bibitem{BR75} R.~ Balescu, {\it Equilibrium and nonequilibrium statistical
   mechanics} (Wiley, New York, 1975), Vol.~1.

 \bibitem{TR38} R.C.~Tolman, {\it The principles of ctatistical mechanics}
  (Oxford at the Clarendon Press, 1938).

\bibitem{RJ58} J.S.\ Rowlinson,{\it The properties of real gases},
  Encyclopedia of Physics, Vol. 3/12 (Springer-Verlag, Academic Edition, Berlin, 1958),
  ISBN : 978-3-642-45894-1.

\bibitem{TK66} K.B.~Tolpygo, {\it Thermodynamics
and Statistical Physics}, (Kiev University, Kiev, 1966) (in Russian).

\bibitem{IA71} A.~Isihara, ``Statistical Physics'' (Academic Press, New York, 1971).
  
\bibitem{ZM02} D.~Zubarev, V.~Morozov, and G.~R\'opke, {\it
  Statistical Mechanics of Nobnequilibrium Processes}, Vol. 1
  (Moscow, Fizmatlit, 2002)(in Russian).

\bibitem{VGS-17}
  V. Vovchenko, M. I. Gorenstein, and H. Stoecker,
   Phys. Rev. Lett. 118, 182301 (2017).   
  
\bibitem{VJGS-18}   V. Vovchenko, L. Jiang, M. I. Gorenstein, and H.
  Stoecker, Phys. Rev. C98, 024910 (2018).

\bibitem{BG-08}  V.V. Begun and M.I. Gorenstein,
Phys. Rev. C {\bf 77}, 064903 (2008).
  
\bibitem{G} W. Greiner, L. Neise, and H. St\"ocker, Thermodynamics and Statistical Mechanics, 1995 Springer-
  Verlag New York, Inc.

  \bibitem{Grad-Ryzhik} I.S.~Gradstein and I.M.~Ryzhik, {\it Tables of Integrals,
  Series and Products} (Moscow, Fizmatlit, 4th edition, 1963).

\bibitem{Li}
A.P. Prudnikov, Yu.A. Brychkov, and O.I. Marichev, Integrals and Series, (Moscow, Nauka, 1986).
 
\bibitem{brack} M.~Brack, C.~Guet, and H-B.~Hakanson, Phys. Rep.~{\bf 123},
  275 (1085).

\bibitem{nm}
  H. A. Bethe, Ann. Rev. Nucl. Part. Sci. {\bf 21}, 93 (1971).

 \bibitem{VOV-17}  V. Vovchenko, Phys. Rev. C 96, 015206 (2017). 



\bibitem{mafm}
 A. G. Magner, K. Arita, S. N. Fedotkin, and K. Matsuyanagi,
 Prog. Theor. Phys. {\bf 108} (2002) 853.

 \bibitem{MY11} 
A.G.\ Magner, Y.S.\ Yatsyshyn, K. Arita, and M.\ Brack,  {\it Phys. At. Nucl. }
{\bf 74}, 1445 (2011).

\bibitem{gutzpr}
 M. C. Gutzwiller,
 {\it Chaos in Classical and Quantum Mechanics}
 (Springer-Verlag, New York, 1990).

\bibitem{gutzbook}
 M. C. Gutzwiller,
 {\it Chaos in Classical and Quantum Mechanics}
 (Springer-Verlag, New York, 1990).

 \bibitem{strumag}
 V. M. Strutinsky,
 Nukleonika,{\bf 20} (1975) 679; \\
 V. M. Strutinsky and A. G. Magner,
 Sov.\ Phys.\ Part.\ Nucl.,{\bf 7} (1977) 138.


\bibitem{BB03} 
M.\ Brack and R.K.\ Bhaduri, {\it Semiclassical Physics. 
  Frontiers in Physics} No. 96, 2nd ed. (Westview Press, Boulder, CO, 2003).

\bibitem{fed:jvmp}
 M. V. Fedoryuk,
 Sov.\ J.\ of\ Comp.\ Math.\ and\ Math.\ Phys.,{\bf 4} (1964) 671;
 ibid. {\bf 10} (1970) 286.

\bibitem{maf}  A. G. Magner, K. Arita, S. N. Fedotkin,
  Progr. Theor. Phys., {\bf 115} (2006) 523.
  
\bibitem{MS20}  A.G. Magner, A.I. Sanzhur, S.N. Fedotkin, A.I. Levon, and S. Shlomo,
  arXiv:2006.03868v2 [nucl-th] 2020, submitted to the Phys. Lett. B, 2020.

\bibitem{LP04} E.M.~Lifshitz and Pitajevsky, Vol.~9 (Moscow, Fizmatlit, 2004 )
  (Russian).
 
  \bibitem{AV-15} D.~Anchishkin and V.~Vovchenko,
    J.\ Phys.\ G {\bf 42}, 105102 (2015). 

\end{thebibliography}
\end{document}